\documentclass{emulateapj}
\usepackage{graphics}
\usepackage{enumitem}
\newcommand{\chandra}{\textit{Chandra }}

\begin{document}

%\received{}
%\revised{}
%\accepted{}

\shortauthors{TEMIM ET AL.}

\shorttitle{DEEP CHANDRA OBSERVATION OF G327.1-1.1} 

\title{Late-time Evolution of Composite Supernova Remnants: \\ Deep \textit{Chandra} observations and Hydrodynamical modeling of a crushed pulsar wind nebula in SNR G327.1-1.1}

%\title{Late Evolution of Composite Supernova Remnants: Deep \textit{Chandra} observations and Hydrodynamical modeling of a crushed pulsar wind nebula in SNR G327.1-1.1}

%\title{Evolution of the composite supernova remnant G327.1-1.1: Deep \textit{Chandra} observations and Hydrodynamical modeling of a crushed pulsar wind nebula}

\author{TEA TEMIM\altaffilmark{1,2}, PATRICK SLANE\altaffilmark{3}, CHRISTOPHER KOLB\altaffilmark{4}, JOHN BLONDIN\altaffilmark{4}, JOHN P. HUGHES\altaffilmark{5} AND NICCOL\'O BUCCIANTINI\altaffilmark{6}}

\altaffiltext{1}{Observational Cosmology Lab, Code 665, NASA Goddard Space Flight Center, Greenbelt, MD 20771, USA}
\altaffiltext{2}{CRESST, University of Maryland-College Park, College Park, MD 20742, USA}
\altaffiltext{3}{Harvard-Smithsonian Center for Astrophysics, 60 Garden Street, Cambridge, MA 02138, USA}
\altaffiltext{4}{North Carolina State University}
\altaffiltext{5}{Rutgers University}
\altaffiltext{6}{INAF Osservatorio Astrofisico di Arcetri}

%\slugcomment{Accepted by ApJ}

\begin{abstract}

In an effort to better understand the evolution of composite supernova remnants (SNRs) and the eventual fate of relativistic particles injected
by their pulsars, we present a multifaceted investigation of the interaction between a pulsar wind nebula (PWN) and its host SNR G327.1-1.1. Our 350~ks \textit{Chandra} X-ray observations of SNR G327.1-1.1 reveal a highly complex morphology; a cometary structure resembling a bow shock, prong-like features extending into large arcs in the SNR interior, and thermal emission from the SNR shell. Spectral analysis of the non-thermal emission offers clues about the origin of the PWN structures, while enhanced abundances in the PWN region provide evidence for mixing of supernova ejecta with PWN material. The overall morphology and spectral properties of the SNR suggest that the PWN has undergone an asymmetric interaction with the SNR reverse shock (RS) that can occur as a result of a density gradient in the ambient medium and/or a moving pulsar that displaces the PWN from the center of the remnant. We present hydrodynamical simulations of G327.1-1.1 that show that its morphology and evolution can be described by a $\sim$ 17,000~yr old composite SNR that expanded into a density gradient with an orientation perpendicular to the pulsar's motion. We also show that the RS/PWN interaction scenario can reproduce the broadband spectrum of the PWN from radio to $\gamma$-ray wavelengths. The analysis and modeling presented in this work have important implications for our general understanding of the structure and evolution of composite SNRs.

\end{abstract}


\section{INTRODUCTION} \label{intro}

Composite supernova remnants (SNRs) are a class of SNRs for which we observe both a synchrotron-emitting pulsar wind nebula (PWN) generated by the wind of a highly magnetized, rapidly rotating pulsar, and a shell from the supernova (SN) blast wave expanding into the ambient interstellar medium (ISM). 
The PWN is characterized by an inner shock where the wind flow is terminated as it enters the expanding nebula, and an outer shock at the SN ejecta interface. Its spectrum is determined by the evolving population of injected particles that spiral in the magnetic field and emit synchrotron radiation. This rotating magnetic field also results in jet and torus structures that have now been observed in a number of systems \citep[review by][]{gaensler06,kargaltsev08}. The interaction of the PWN with its host SNR significantly modifies the relativistic particle population that eventually escapes into the ISM, and it may also give rise to the low-energy particle excess that produces $\gamma$-ray emission through inverse-Compton scattering \citep[e.g.,][]{abdo10,slane10,slane12,temim13a}. This interaction also reveals information about the evolution of the PWN and SNR, the properties of the SN ejecta, progenitor star, and the ambient medium into which the SNR expands. For these reasons, the study of composite SNRs, particularly in the late stages of evolution when the SN reverse shock (RS) interacts with the PWN, is of considerable importance.

%**************************************************
\begin{figure*}
\epsscale{0.55} \plotone{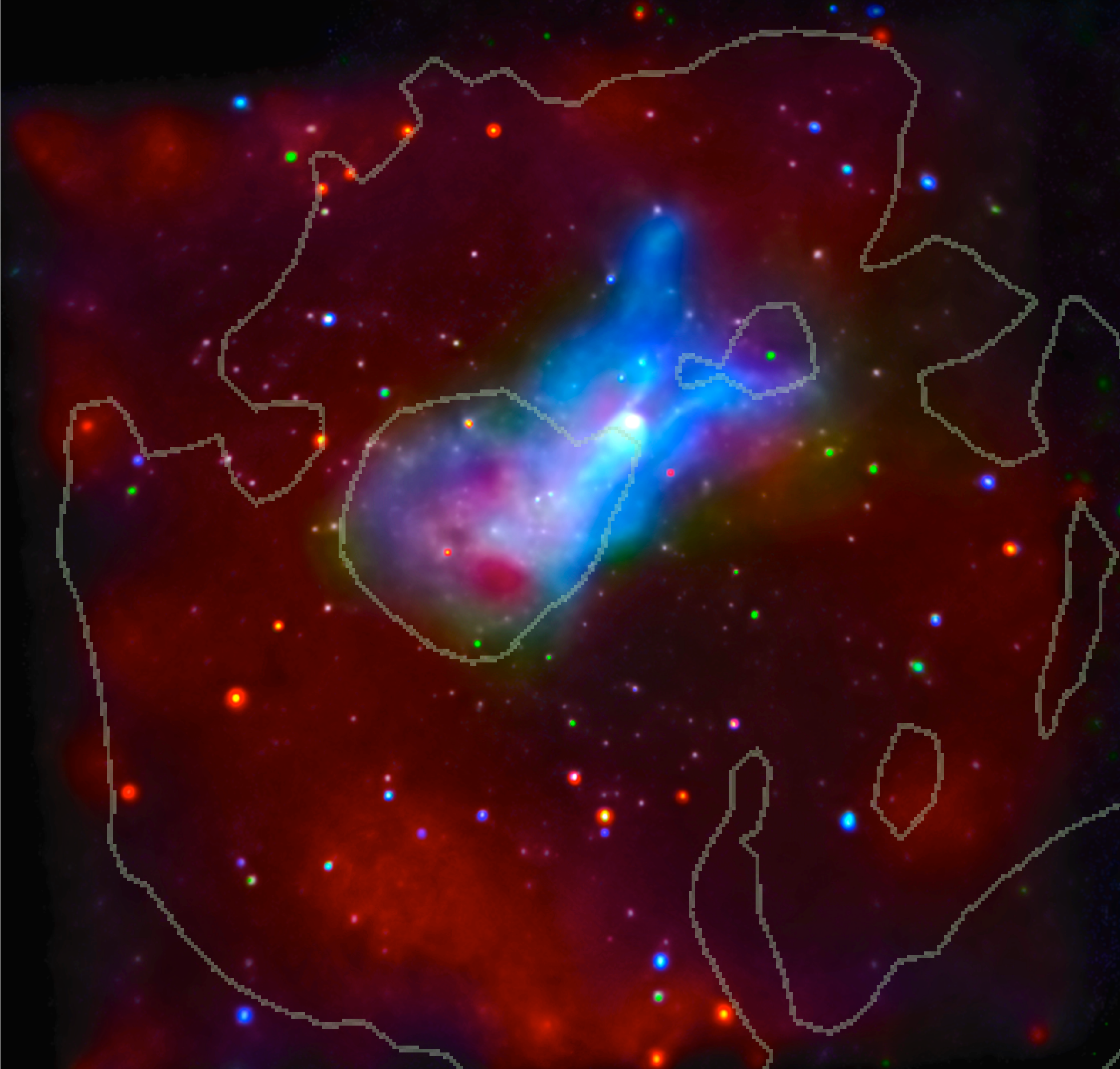}
\epsscale{0.55} \plotone{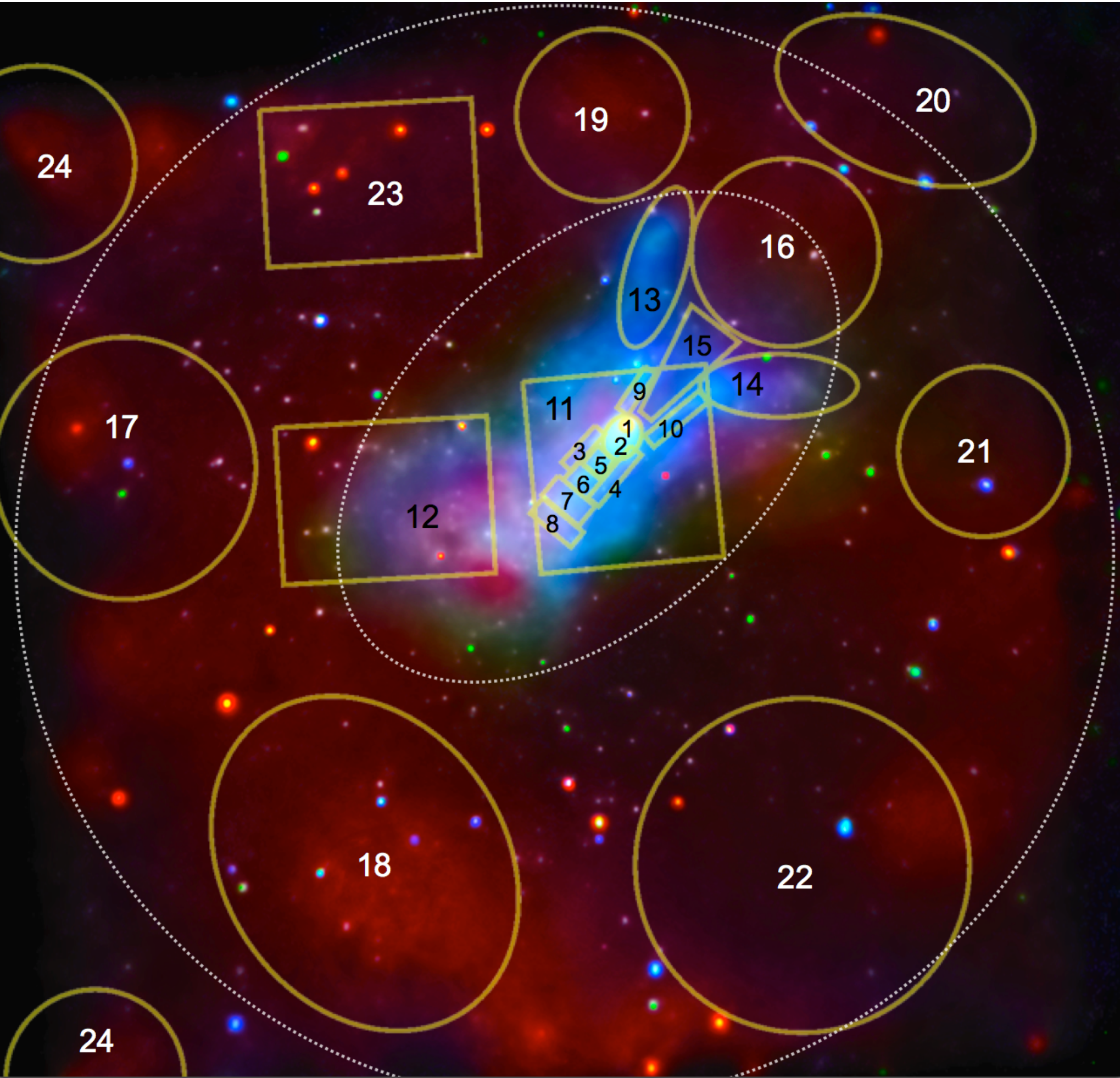}

\caption{\label{3color}A false color \chandra X-ray image of G327.1-1.1 with the 0.75-1.45 keV emission in red, 1.45-2.58 keV emission in green, and 2.58-7.0 keV emission in blue\footnote{http://chandra.harvard.edu/photo/2014/msh11g327/ (NASA/CXC/GSFC/T.Temim et al.)}. The Molonglo Observatory Synthesis Telescope (MOST) 843 MHz radio contours \citep{whiteoak96} are overlaid on the left panel, with the inner contours outlining the radio relic PWN, and the outer counter the SNR shell, approximately 17\arcmin in diameter. The right panel shows the regions used for the spectral extraction; the white dotted regions correspond to those listed in Table~\ref{spec1} and shown in Figure~\ref{specfig1}, while the numbered regions correspond to those listed in Table~\ref{spec2} and shown in Figures~\ref{specfig2} and \ref{specfig3}.}
\end{figure*}
%**************************************************

G327.1-1.1 is  a composite SNR that appears to be in the late stages of evolution, with a PWN that has likely undergone an interaction with the RS \citep{temim09}.
At these late stages of composite SNRs' evolution, the RS propagates back into the cold SN ejecta in the SNR interior and reaches the boundary of the expanding PWN. Since the density of the medium surrounding a core-collapse SN is unlikely to be uniform, the RS typically returns preferentially from the direction of higher ambient density. This, along with any offset of the PWN due to the motion associated with the pulsar itself, results in an asymmetric PWN/RS interaction that disrupts the nebula and gives rise to complex structures and mixing of the SN ejecta with PWN material \citep{blondin01,vanderswaluw04}. The low-energy particles, whose synchrotron losses during the crushing phase have not been extreme, persist as a relic radio nebula, while fresh energetic particles injected from the pulsar begin to reform an X-ray PWN that is displaced from the radio emission due the passage of the reverse shock and/or the pulsar's motion. Such morphology has been observed in composite SNRs, such as G327.1-1.1 and MSH 15-56 \citep{temim09,temim13a}. In cases where the pulsar's motion through the SNR becomes supersonic, the X-ray nebula deforms into a bow shock with a cometary morphology, as observed in the Mouse \citep{gaensler04}.

SNR G327.1-1.1 contains a bright central PWN whose structure is complex in both radio and X-ray bands. It shows evidence for all of the properties described above; an SNR shell, a relic radio nebula resulting from an asymmetric RS interaction, a newly reformed X-ray PWN that appears to be deforming into a bow shock due to the pulsar's rapid motion \citep{temim09}, and TeV $\gamma$-ray emission that spatially coincides with the PWN \citep{acero12}. There is still no evidence for pulsed emission from the putative pulsar that is forming the PWN. The previous 85 ks \textit{XMM-Newton} and 50 ks \chandra observations revealed the PWN's unusual structures, such as the prong-like features extending into what appeared to be a bubble in the SNR interior. While these observations led to the determination of the SNR's overall properties and basic structure, many questions still remain about the nature of this system and its evolution. We still don't understand what causes the PWN's unusual morphology, what is its current evolutionary stage, and how the injected particle population is altered by the reverse shock interaction. In this paper, we present a study of the deep 350~ks \chandra observations of the composite SNR G327.1-1.1, with the goal of understanding its physical properties and evolutionary stage. We also present a 2-dimensional (2D) hydrodynamical (HD) model of G327.1-1.1 that provides insight into the evolution and origin of the morphology of this remnant and composite SNRs in general.

The paper is organized as follows. Sections~\ref{obsv} and \ref{analysis} describe the observations and basic analysis of the data. Section~\ref{morph} discusses the general morphology of the SNR and the PWN structures revealed by the deep \textit{Chandra} observation. Secion~\ref{spectra} describes the spectral fitting results and discusses the spectral properties of the SNR shell and the PWN.  Section~\ref{hdmodel} presents the HD model for the evolution of G327.1-1.1, simulated as an SNR blast-wave expanding in an ISM density gradient and containing a PWN produced by a rapidly moving pulsar. Section~\ref{broadband} presents our modeling of the broadband emission of G327.1-1.1 from radio to $\gamma$-ray wavelengths. Section~\ref{evol} summarizes the results on the evolutionary state of the SNR and some of the outstanding questions. The conclusions are summarized in Section~\ref{conclusions}.

\section{OBSERVATIONS AND DATA REDUCTION} \label{obsv}

A deep observation of the SNR G327.1-1.1 was carried out with the Advanced CCD Imaging Spectrometer, ACIS-I, on board the \textit{Chandra X-ray Observatory} for a total exposure time of 350~ks. The observations were taken in the VFAINT mode, on 2012 May 22, 24 and 27, under the observation IDs 13767, 13768, and 14430, and correspond to exposure times of 143, 170, and 37~ks, respectively. The standard data reduction and cleaning were performed using the \textit{chandra\_repro} script in CIAO version 4.5, resulting in a total clean exposure time of 337.5 ks.

The merged, exposure corrected and adaptively smoothed false color image of G327.1-1.1 is shown in Figure \ref{3color}, with the 0.75-1.45 keV shown in red, 1.45-2.58 keV in green, and 2.58-7.0 keV emission in blue. The hard X-ray emission shown in blue originates from the PWN material, while the soft X-ray emission in red shows the spatial distribution of the thermal emission. The overlaid radio contours from the Molonglo Observatory Synthesis Telescope (MOST) 843 MHz observation \citep{whiteoak96} show the outline of the SNR shell (17\arcmin\ in diameter) and the inner relic PWN. The right panel of Figure \ref{3color} shows the regions used for the spectral extraction. Spectra were extracted and response files generated for each observation ID individually, using the CIAO \textit{specextract} script, producing a set of three spectra and associated files for each spectral region. Since the observations were carried out close in time and have the same roll angle, the spectra and response files were merged to produce a single spectrum for each extraction region.

%**************************************************

\begin{figure}

\center

\epsscale{1.15} \plotone{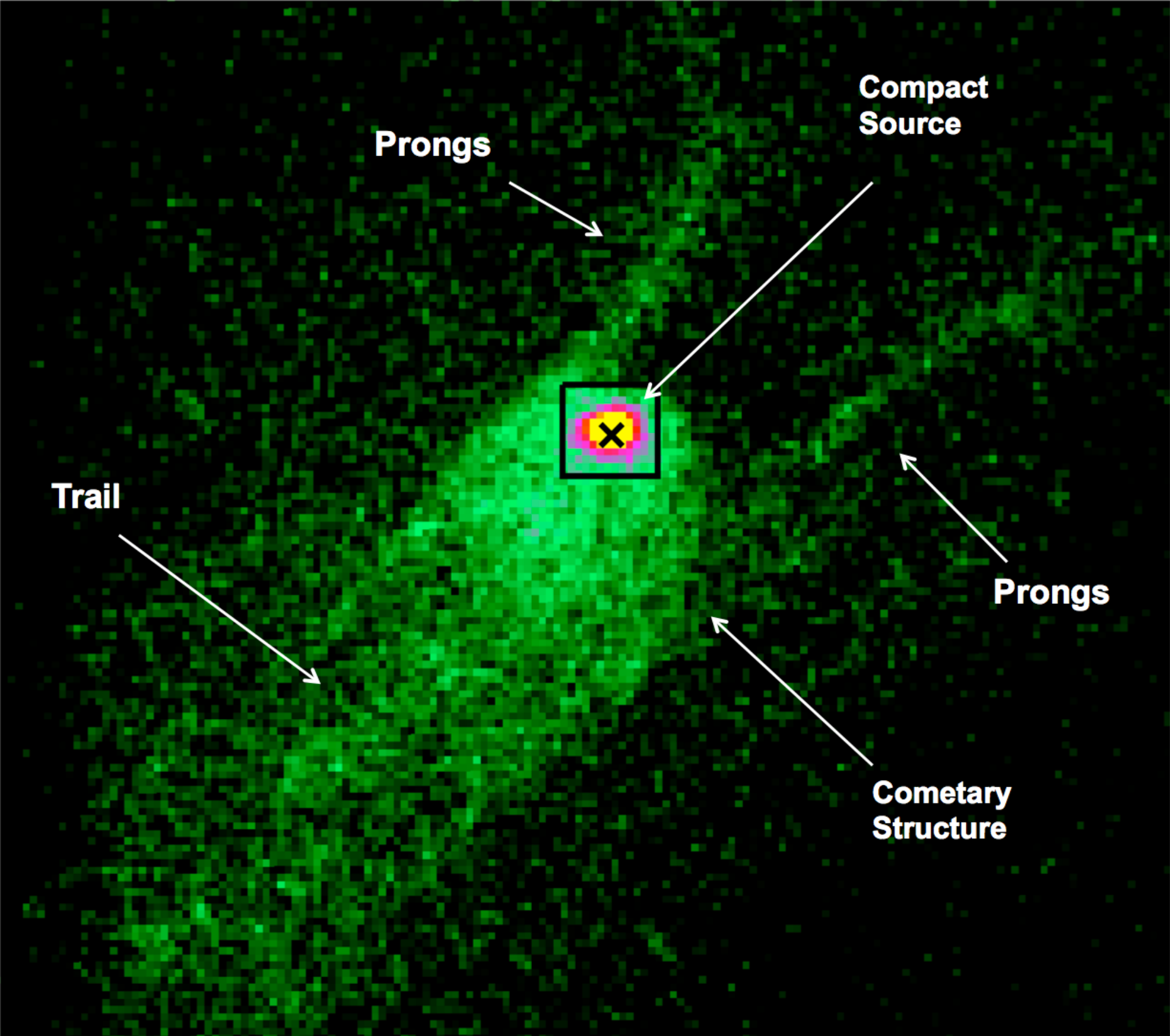}
\epsscale{1.2} \plotone{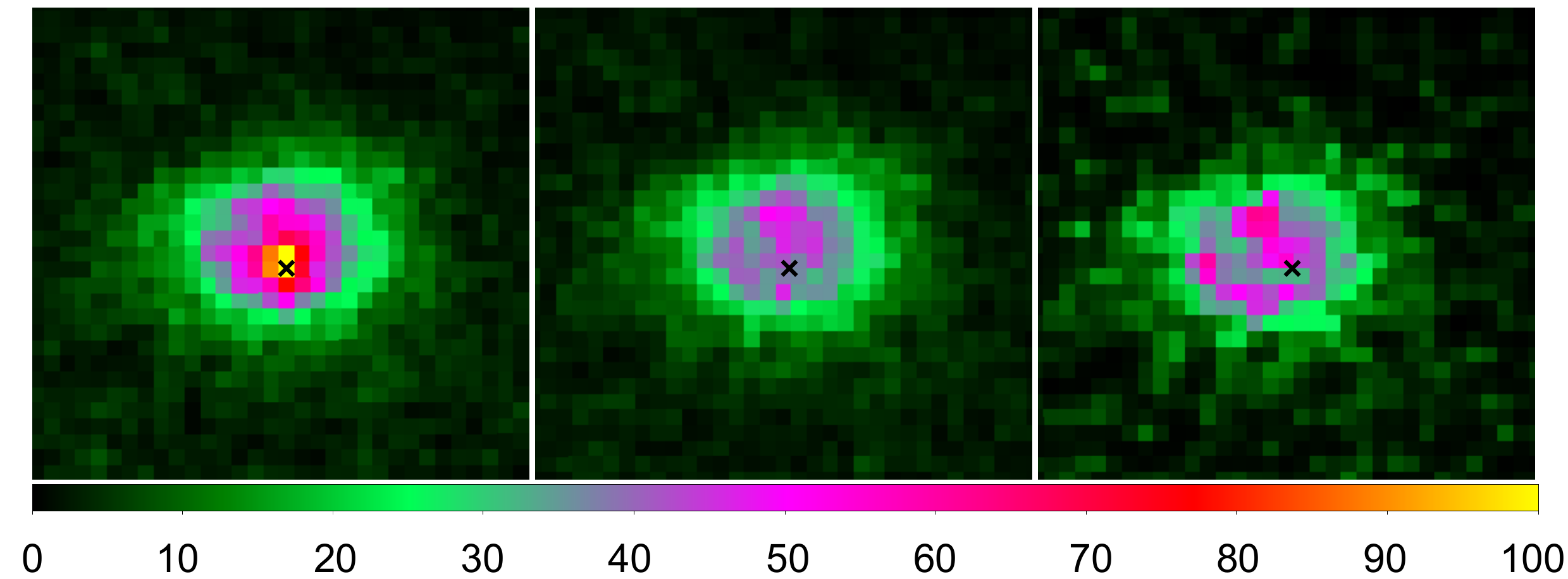}

\caption{\label{pwn} Top panel shows the unsmoothed \chandra ACIS image of the PWN region (2\farcm6 across), corresponding to region 11 in Figure~\ref{3color} (right). 
The black box is 20\arcsec\ in size and indicates the zoomed-in region of the compact source shown in the bottom panel. The bottom panel's left image shows the unbinned image of the compact source, the middle panel shows the residual
image after the subtraction of a point source at
the location of the black ``x", and the right panel shows the same residual deconvolved
by the task \textit{arestore}. The scale bar is the units of counts.}

\end{figure}
%**************************************************

\section{ANALYSIS}\label{analysis}

In order to study the properties of the various structures observed in the PWN and the spatial variations in the thermal emission across the SNR shell, we fitted spectra from 23 different source regions, shown in Figure \ref{3color}.
Since the SNR shell covers most of the field of view, we used the ACIS blank-sky background files adopted to our observation (http://cxc.harvard.edu/ciao/threads/acisbackground/) to extract background spectra from each extraction region in Figure \ref{3color}. We used the high energy data from 10-12 keV to compare the particle background in each ACIS-I chip in our observations to the particle background in the blank-sky data. We found that the particle background in the blank-sky data was $\approx$ 18\% higher, and we therefore scaled the exposure time of the blank-sky data accordingly. We extracted backgrounds from two small regions in our observation of G327.1-1.1 that were free of SNR emission (regions marked ``24" in Figure \ref{3color}), in order to compare them to the corresponding scaled blank-sky background spectra. We found that the scaled blank-sky spectra described the background very well, except for a residual thermal excess between 1-2 keV in the background of our observation. Since the blank-sky data only account for background emission for high Galactic latitudes, the presence of an additional thermal excess in the vicinity of G327.1-1.1, which is located in the Galactic plane, is not surprising. We modeled this excess simultaneously in the two background regions in Figure \ref{3color} and found that it is well described by an \rm{XSVNEI} model with an absorbing column density $N_H=1.07\times10^{22}\: \rm cm^{-2}$ and a temperature of 0.68 keV. We note that these parameters are quite different than those for the thermal emission associated with the SNR.

\begin{figure*}
\epsscale{0.40} \plotone{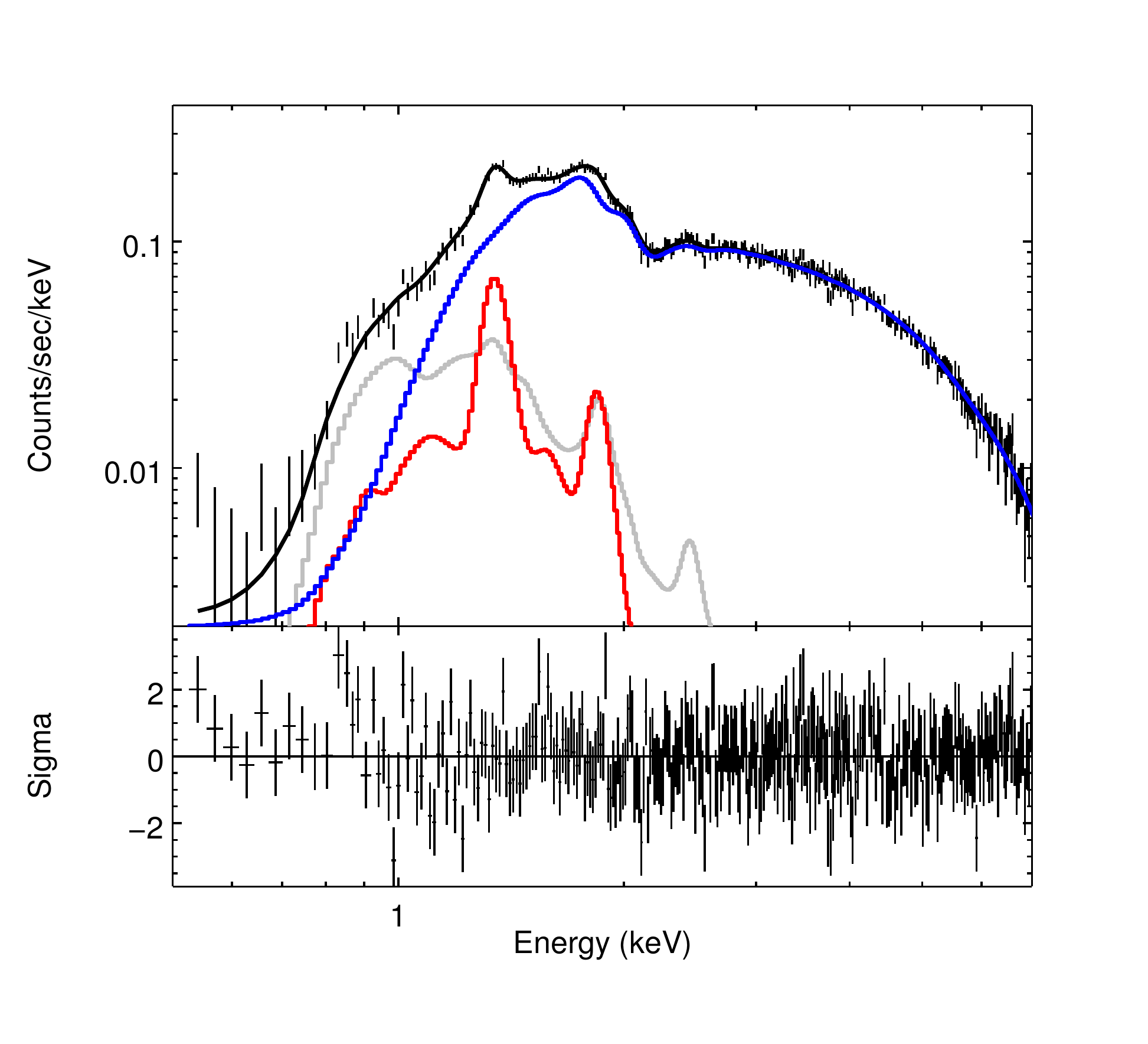}
\hspace{-8mm}
\epsscale{0.40} \plotone{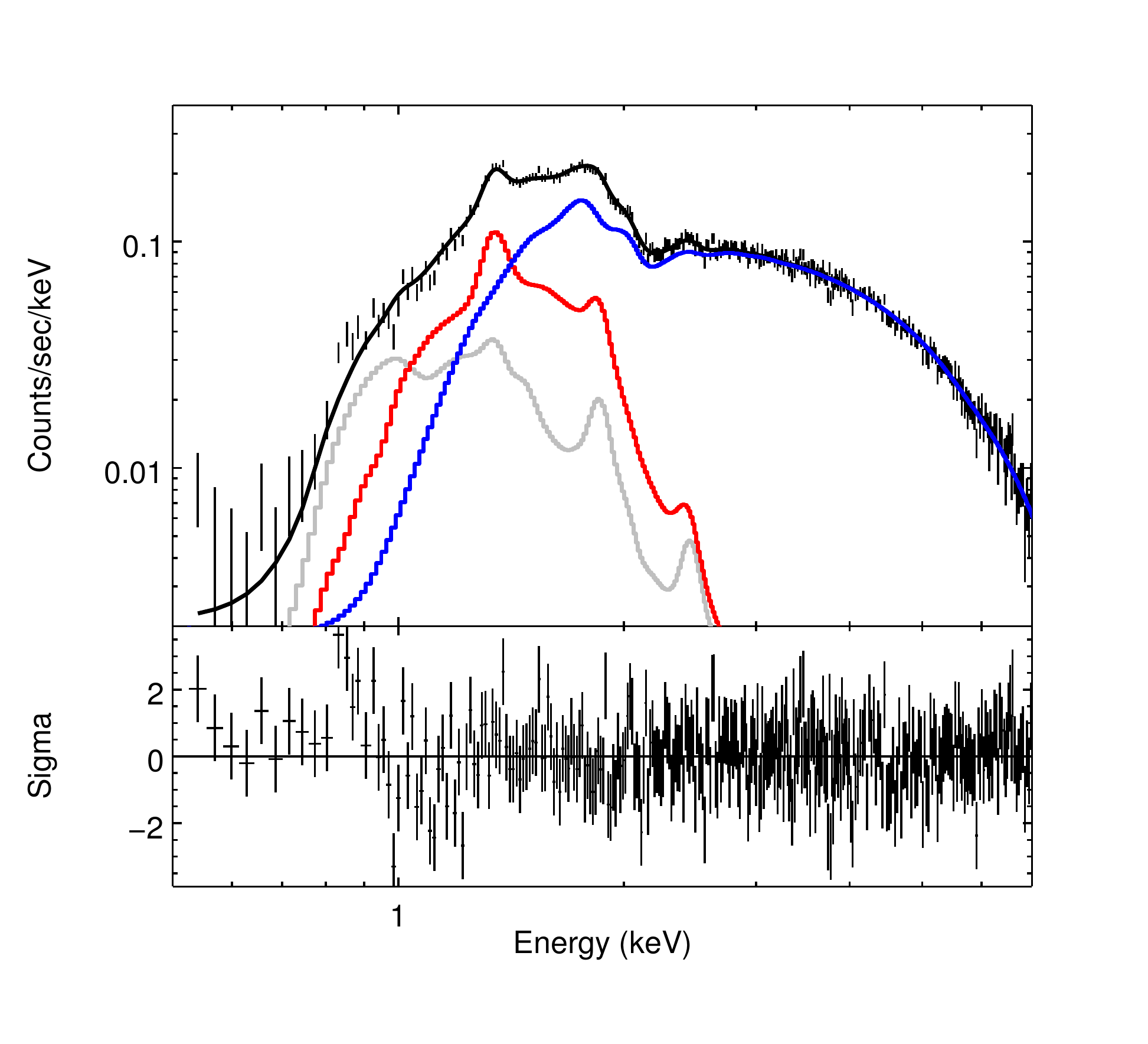}
\hspace{-8mm}
\epsscale{0.41} \plotone{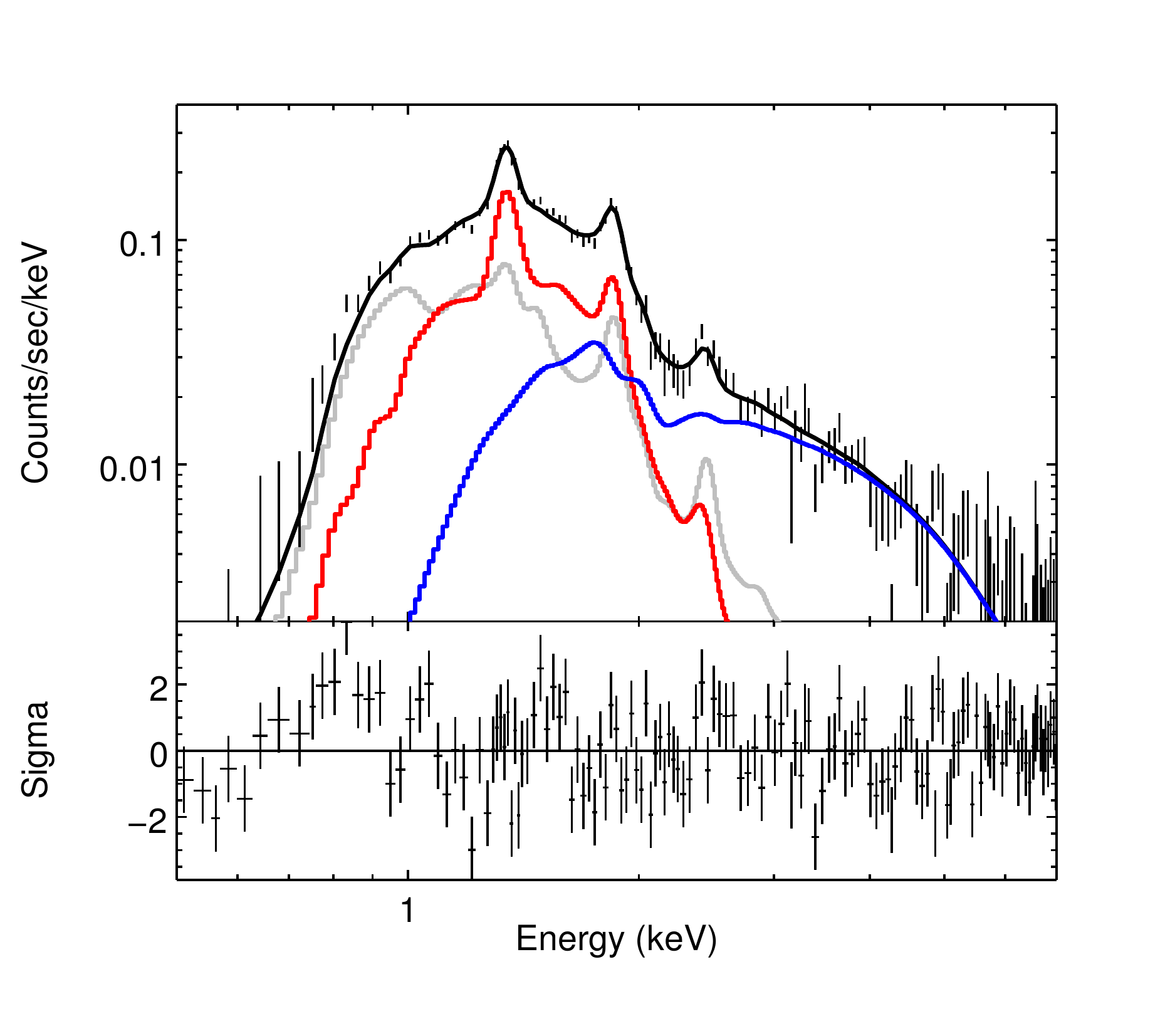}

\caption{\label{specfig1}Best-fit models to the spectra for the whole PWN and SNR shell regions, the white dotted regions in the right panel of Figure~\ref{3color}, where the extraction region for the shell is the annulus that excludes the inner PWN region. For all three panels, the power-law component is shown in blue, thermal component in red, and residual background in gray (see Section~\ref{analysis}).
The left and middle panels show the spectrum and best-fit models for the PWN region, which is equally well fitted by a thermal model with enhanced and solar abundances, respectively. The right panel shows the spectrum and best-fit model for the SNR shell. The values for the best-fit parameters are listed in Table \ref{spec1}.}

\end{figure*}

\begin{figure*}
\epsscale{0.4} \plotone{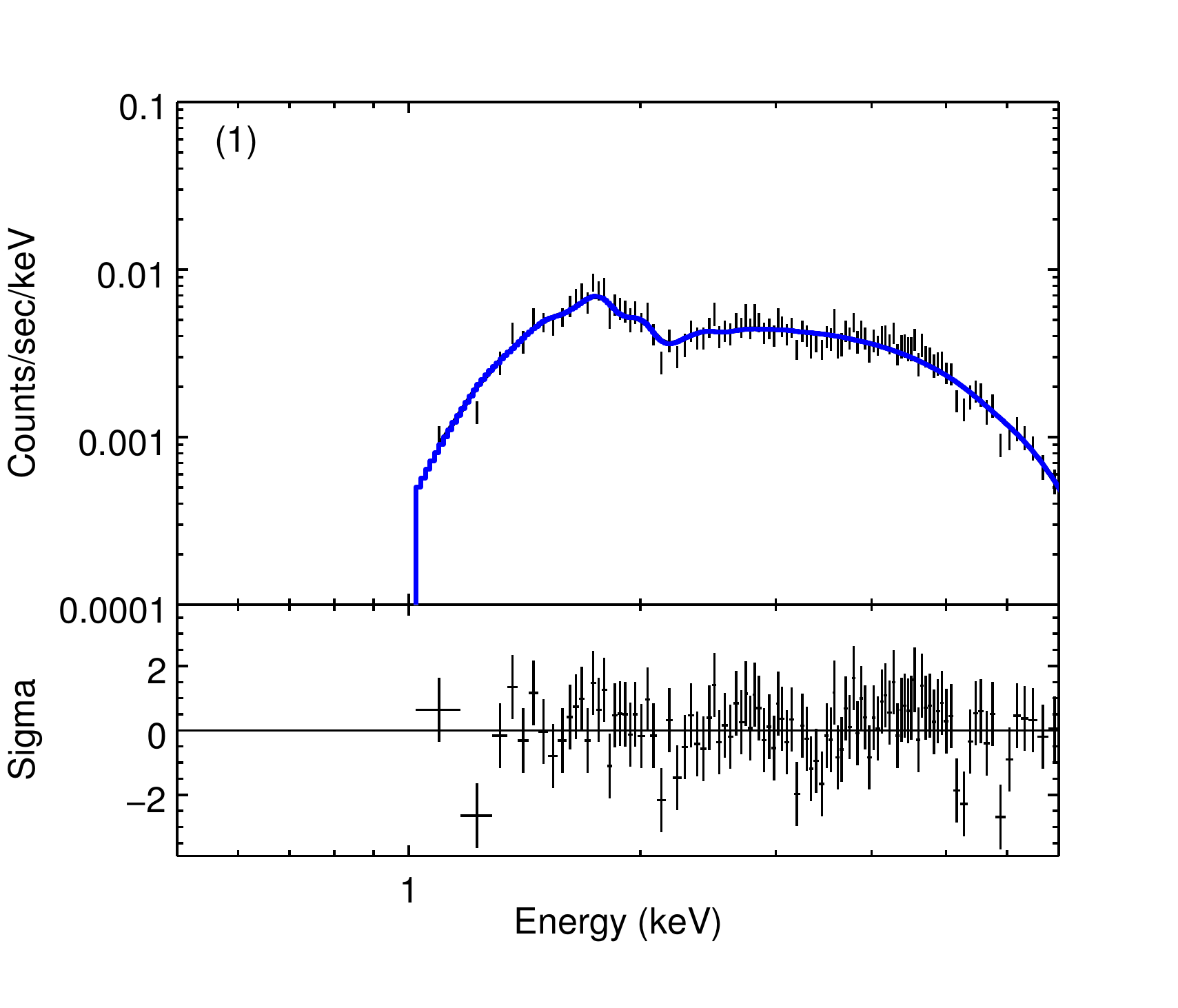}
\hspace{-6.5mm}
\epsscale{0.4} \plotone{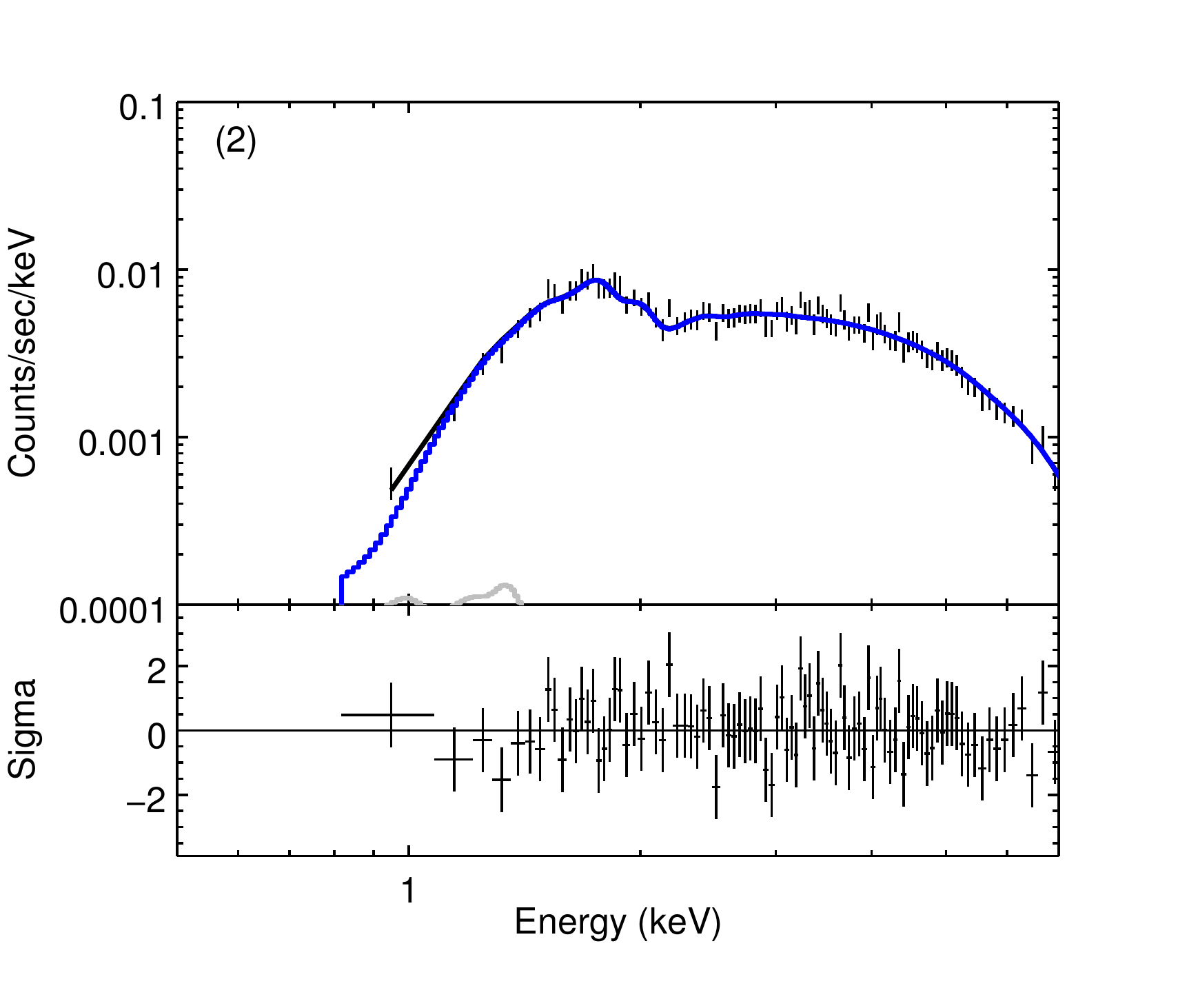}
\hspace{-6.5mm}
\epsscale{0.405} \plotone{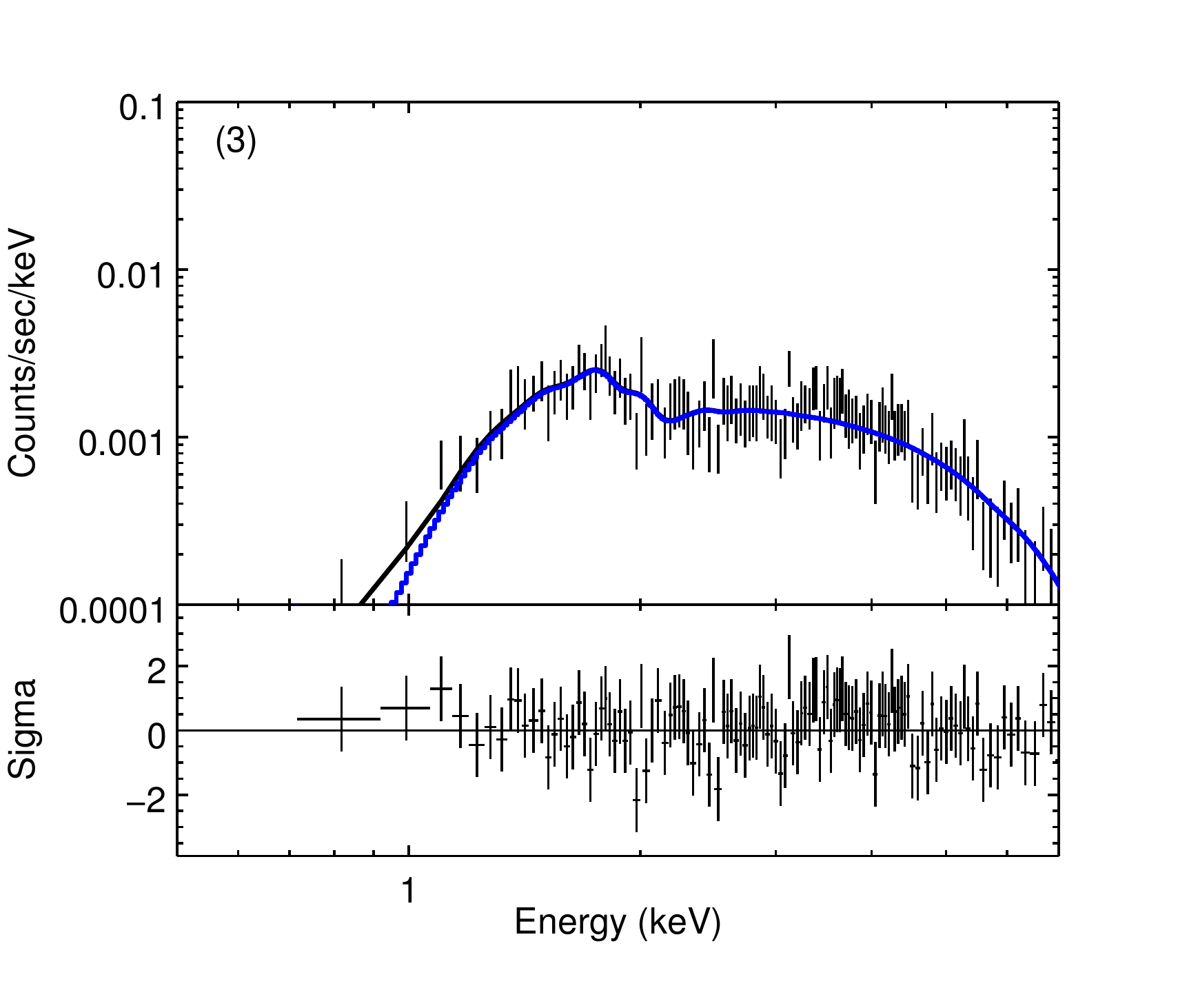}
\vspace{-4mm}
\epsscale{0.4} \plotone{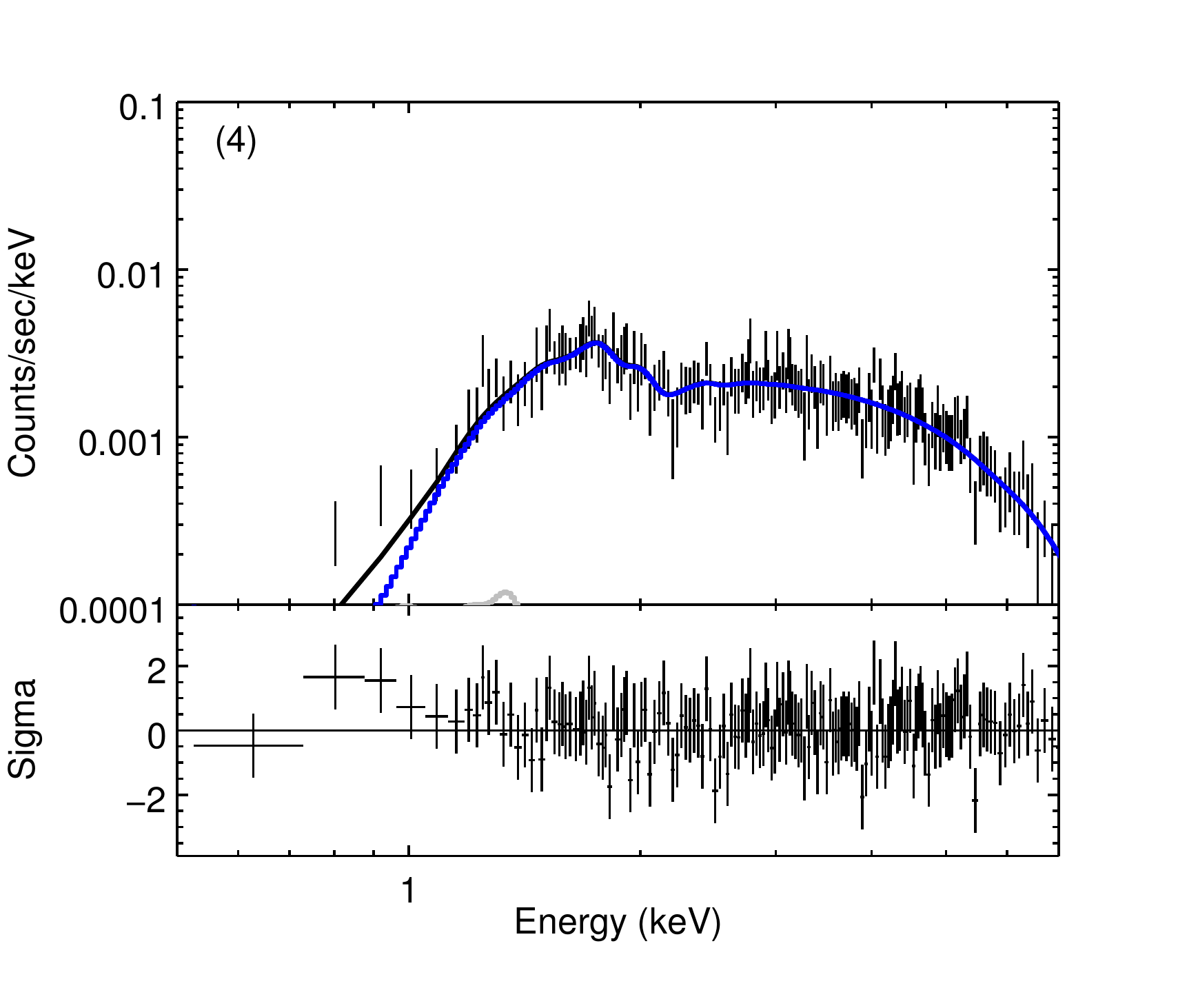}
\hspace{-6.5mm}
\epsscale{0.4} \plotone{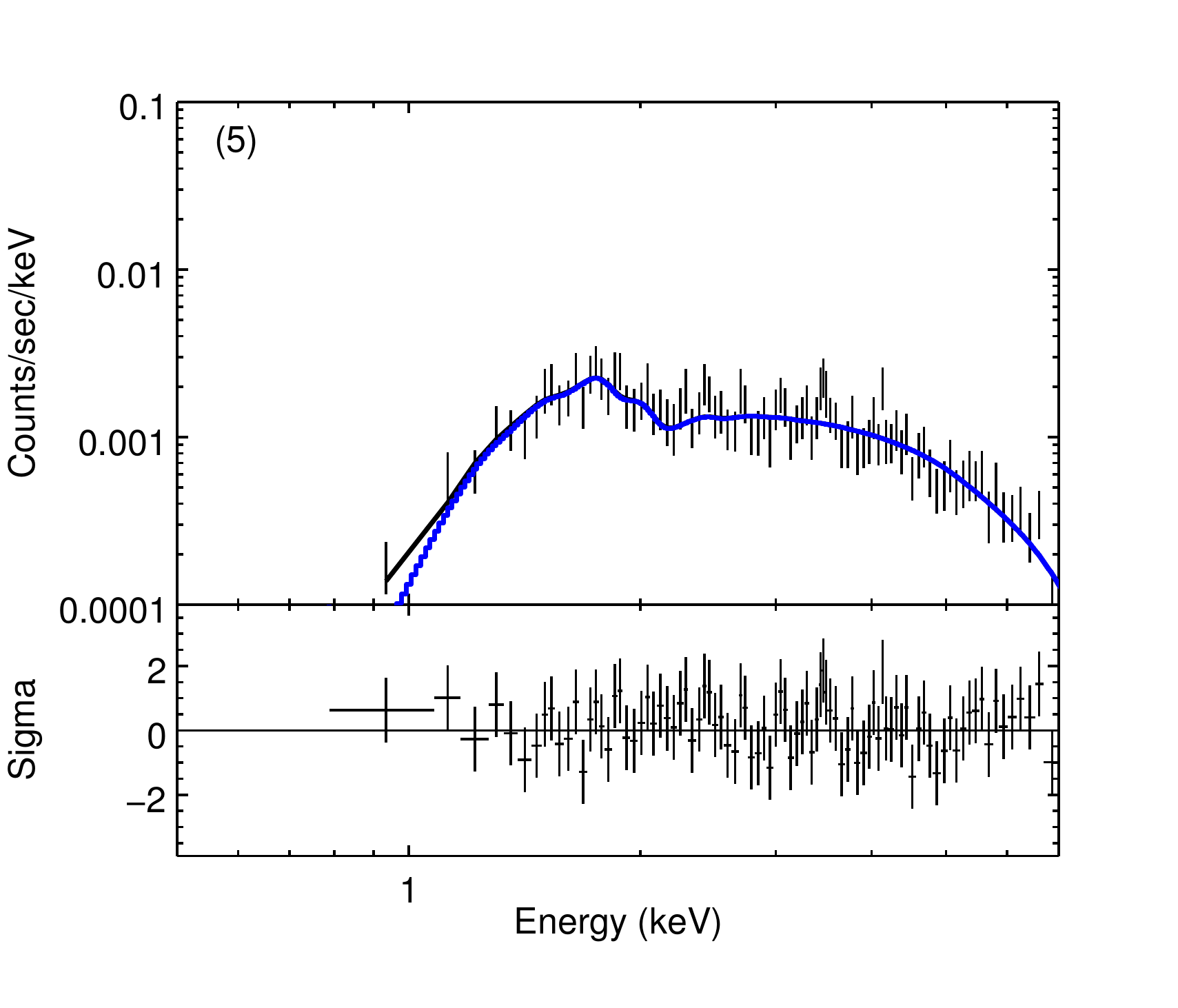}
\hspace{-6.5mm}
\epsscale{0.4} \plotone{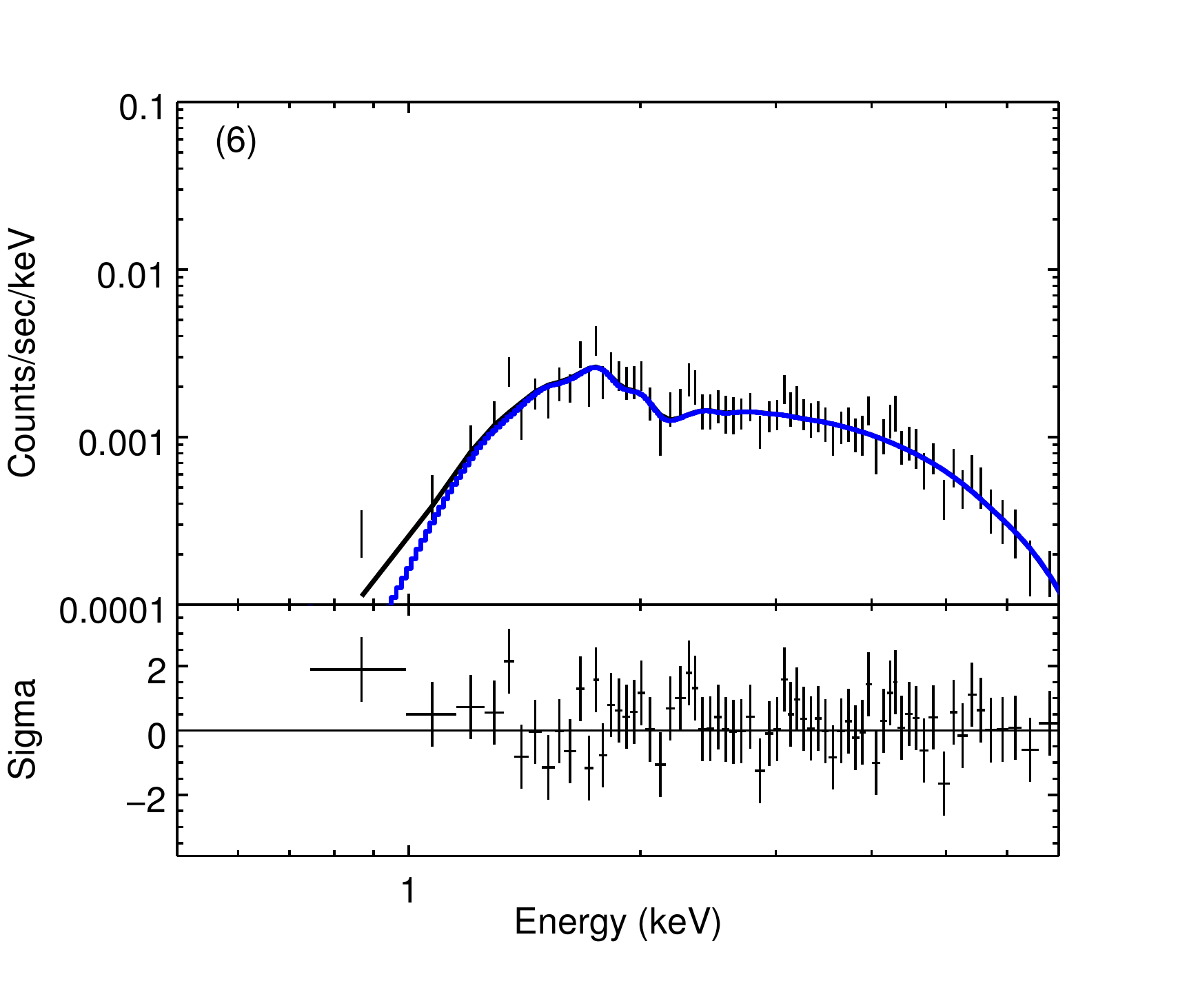}
\vspace{-4mm}
\epsscale{0.4} \plotone{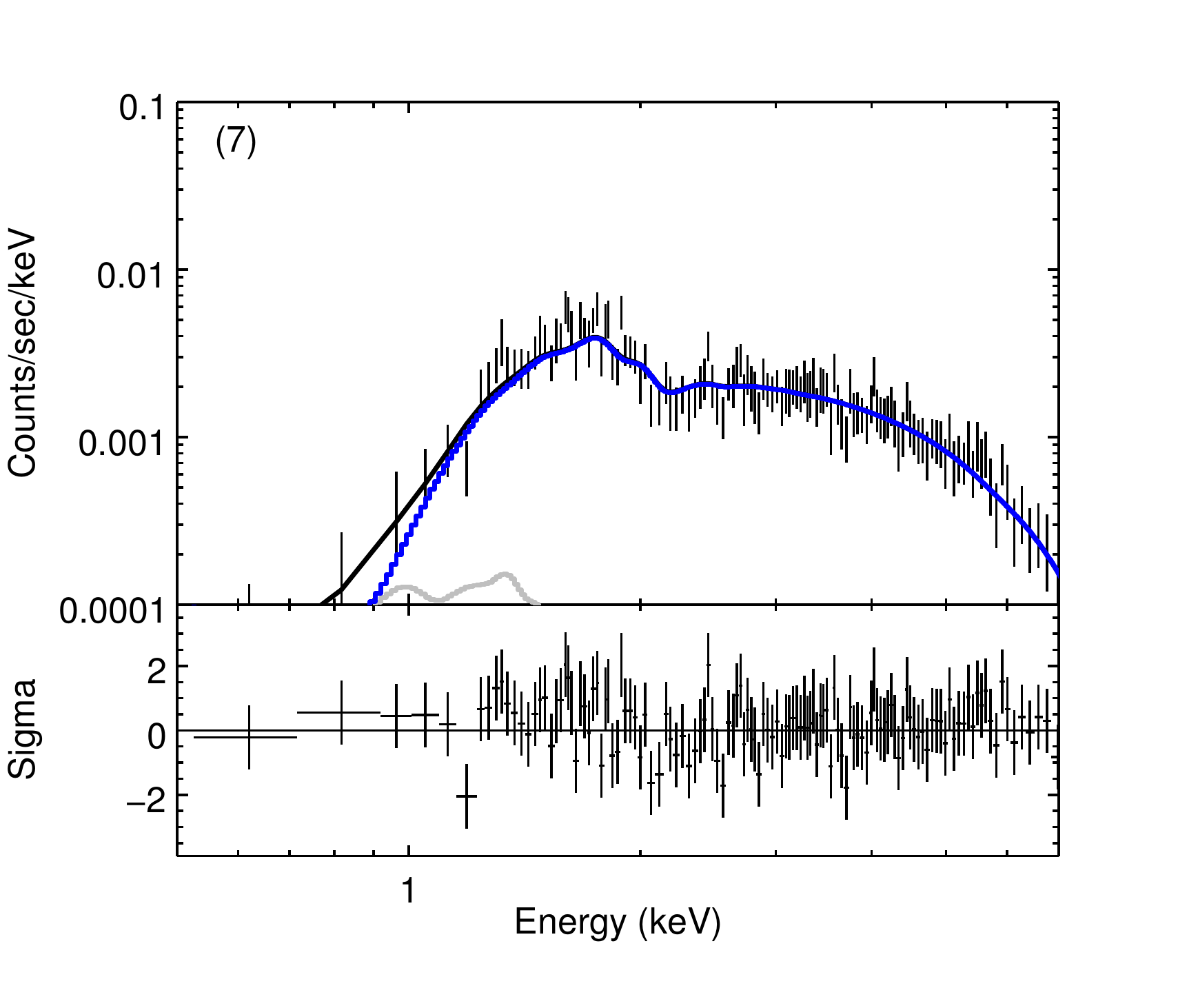}
\hspace{-6.5mm}
\epsscale{0.4} \plotone{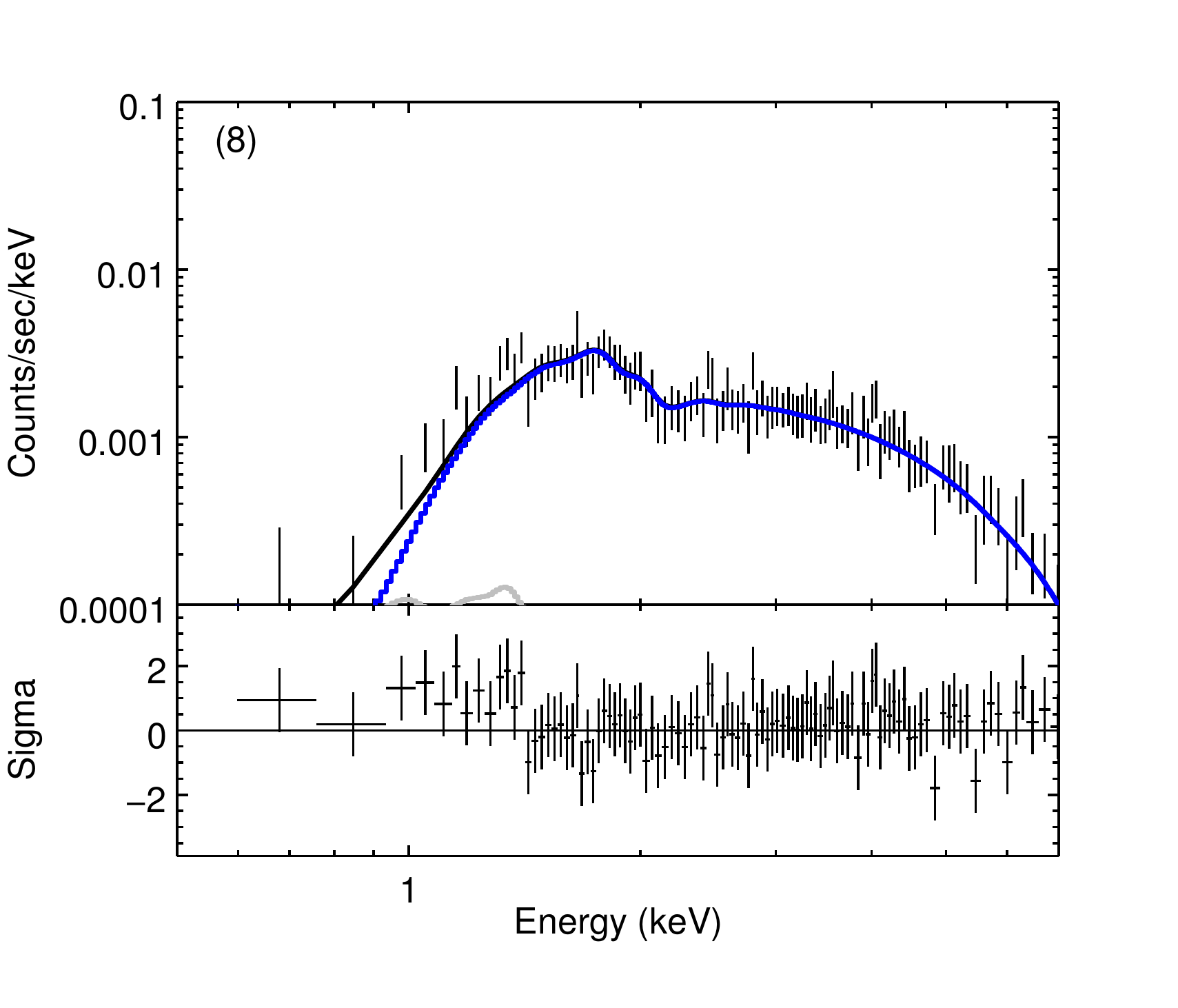}
\hspace{-6.5mm}
\epsscale{0.4} \plotone{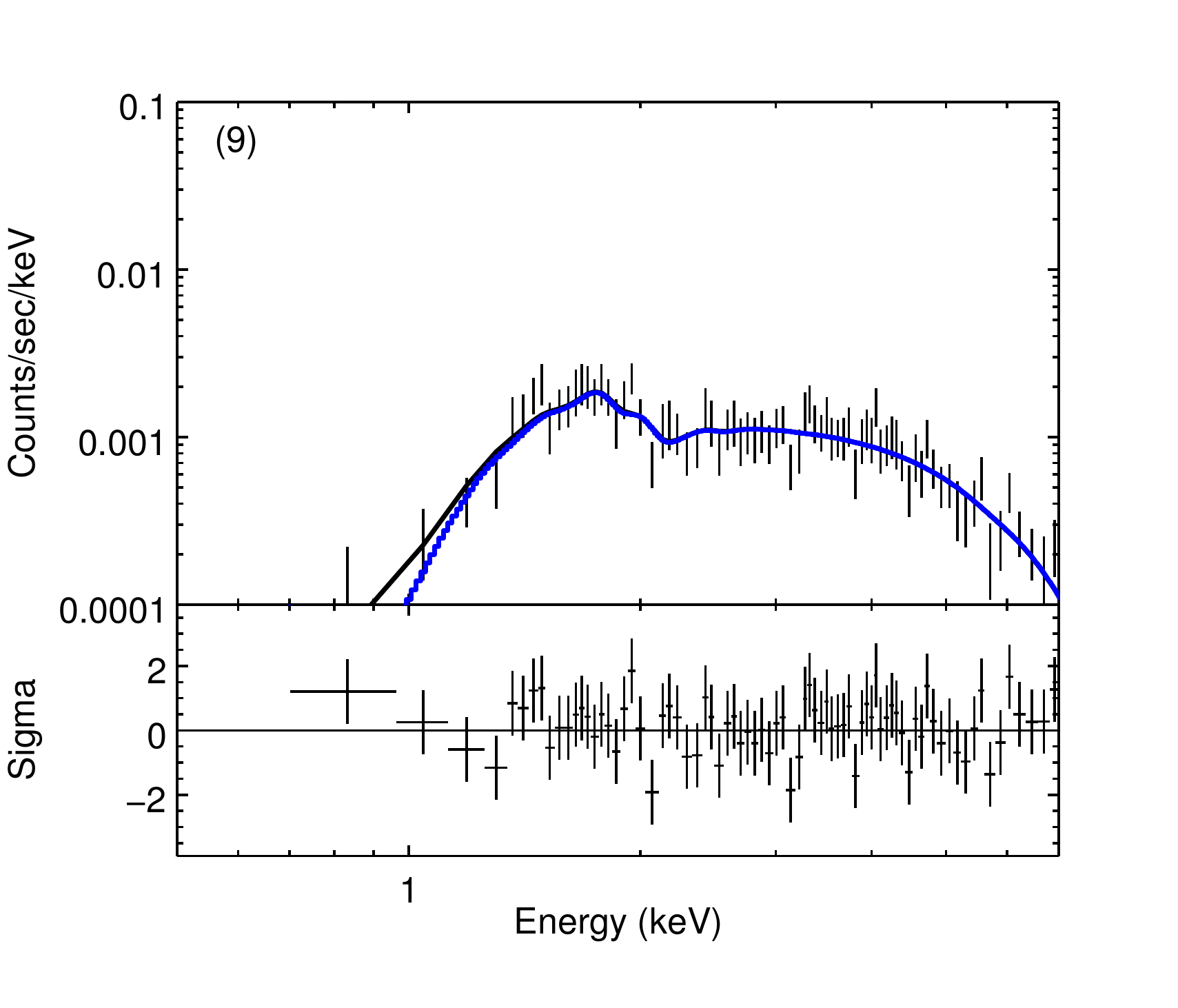}
\vspace{-4mm}
\epsscale{0.4} \plotone{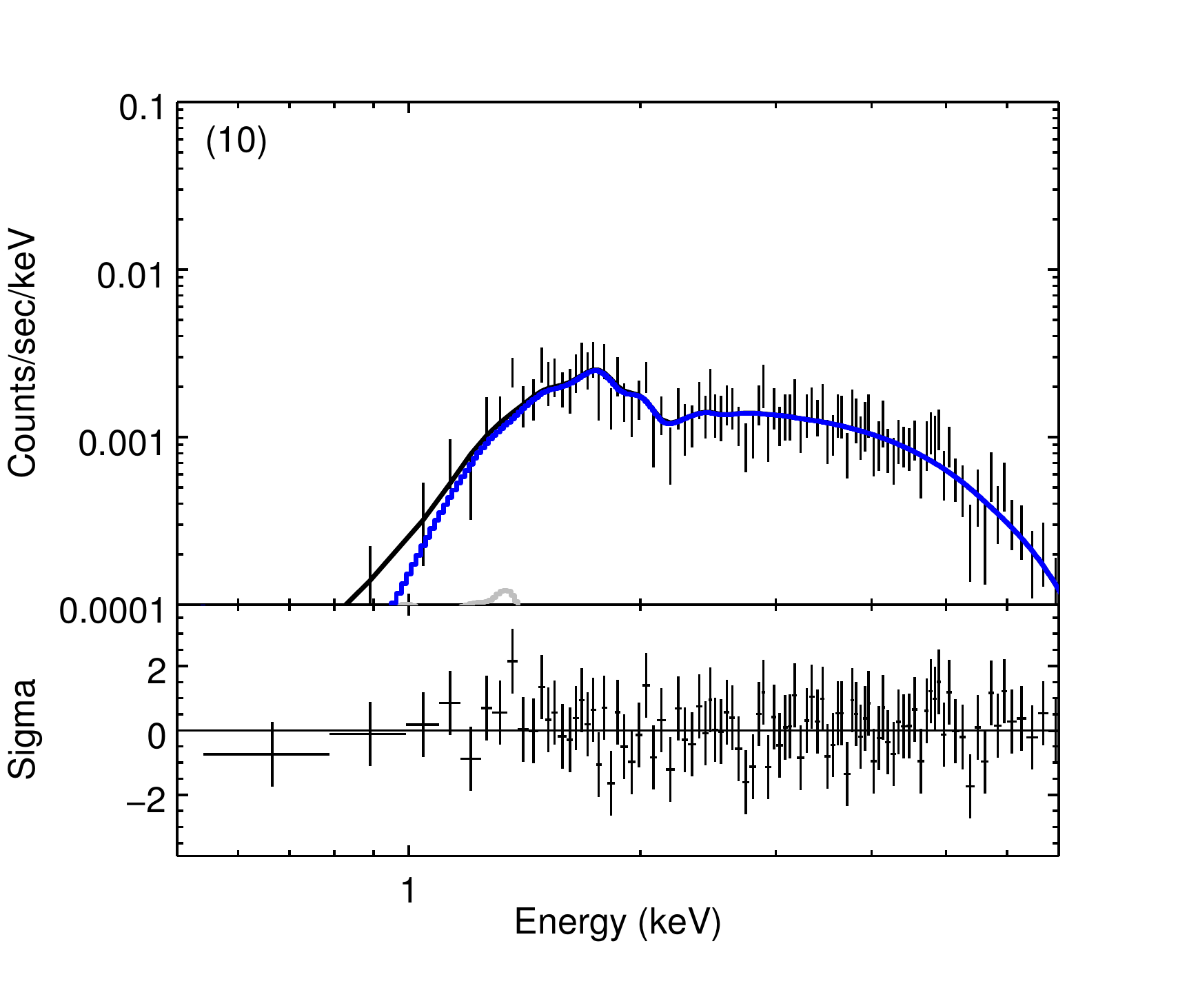}
\hspace{-6.5mm}
\epsscale{0.4} \plotone{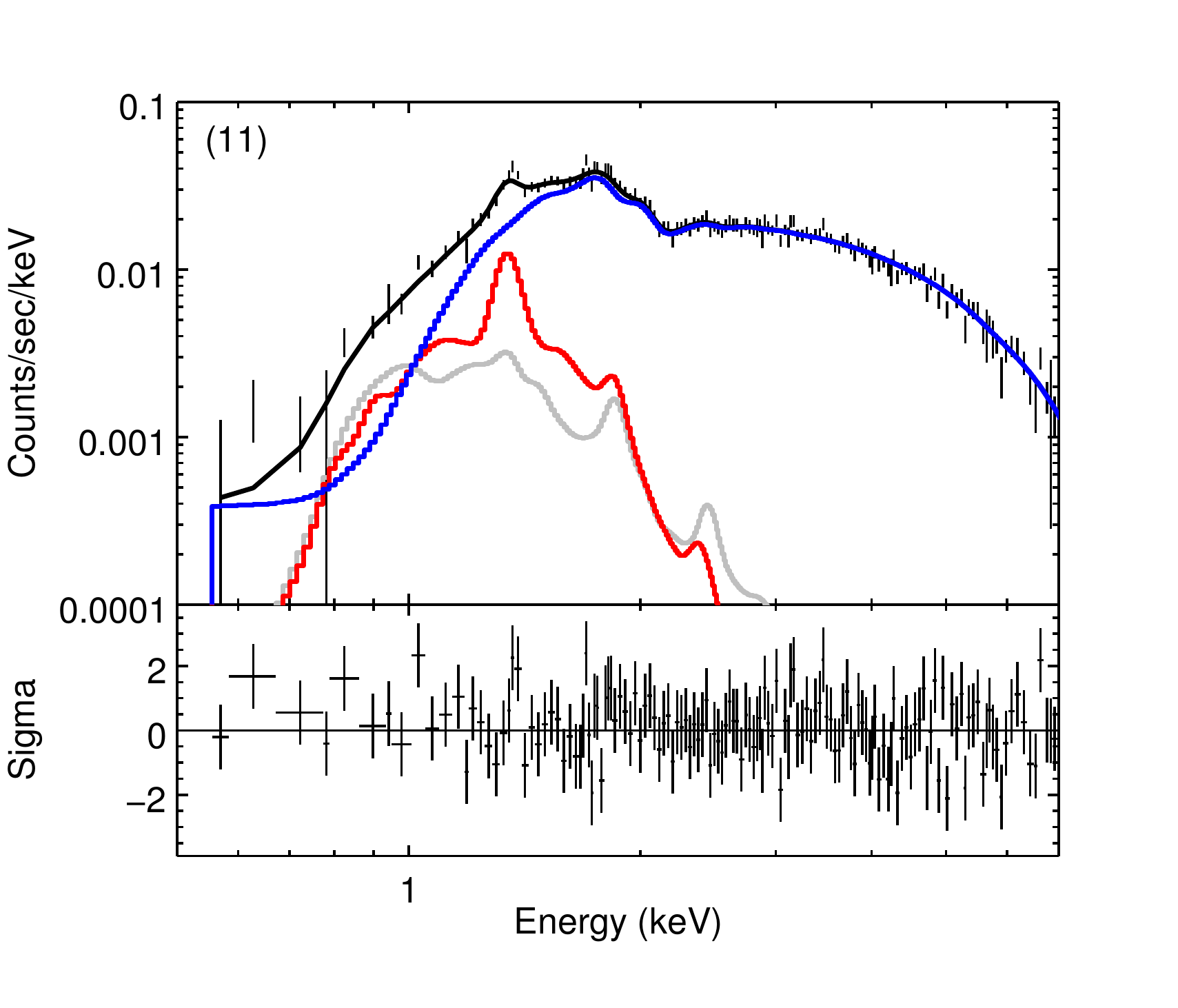}
\hspace{-6.5mm}
\epsscale{0.4} \plotone{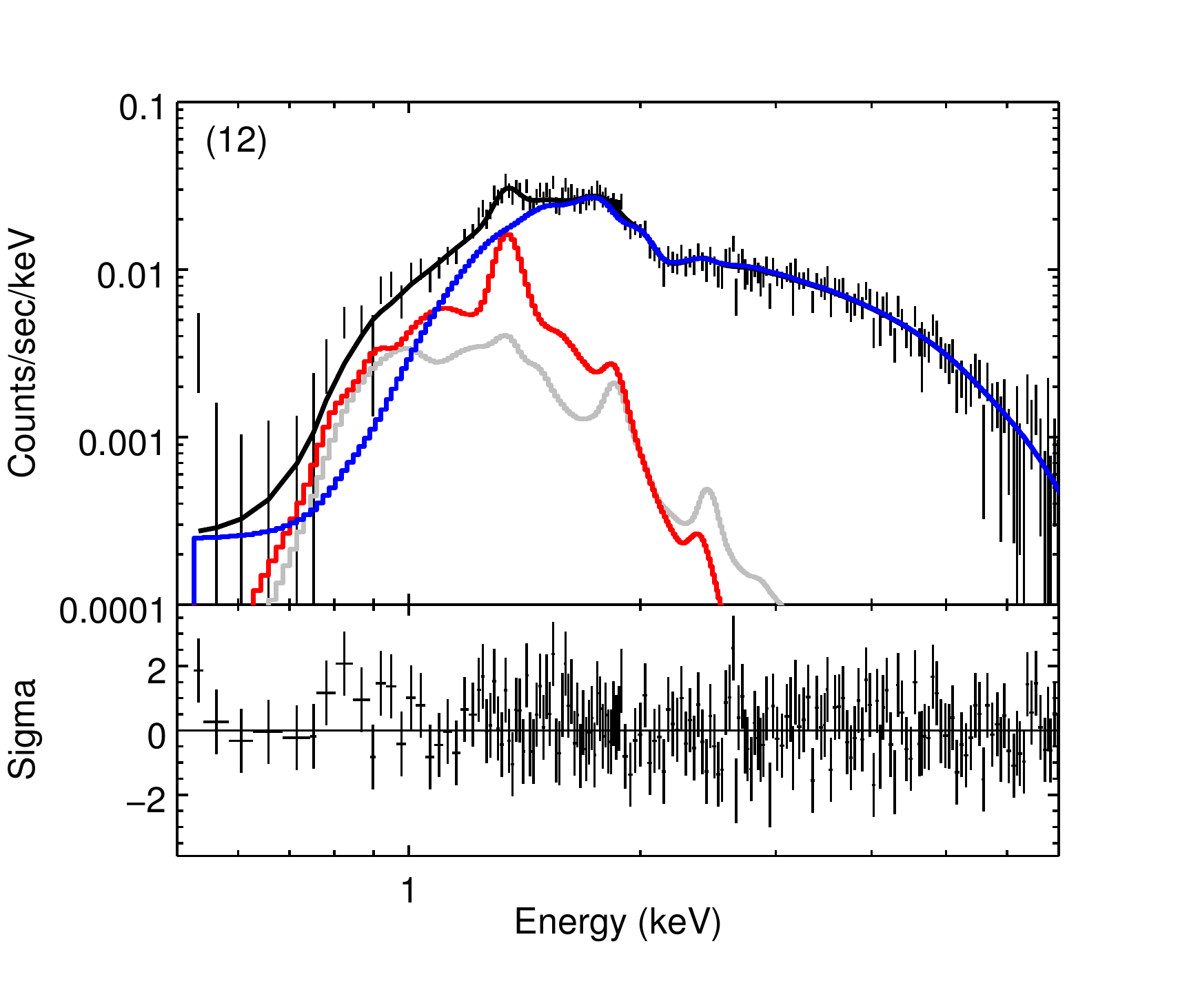}

\caption{\label{specfig2}Best-fit models and residuals for the spectra from regions 1--12 in Figure~\ref{3color} and listed in Table~\ref{spec2}, where the numbers in the upper lefthand corners of the panels correspond to the region numbers. For all panels, the power-law component is shown in blue, thermal component in red, and residual background after blank-sky subtraction in gray (see Section~\ref{analysis}).}

\end{figure*}

\begin{figure*}
\epsscale{0.4} \plotone{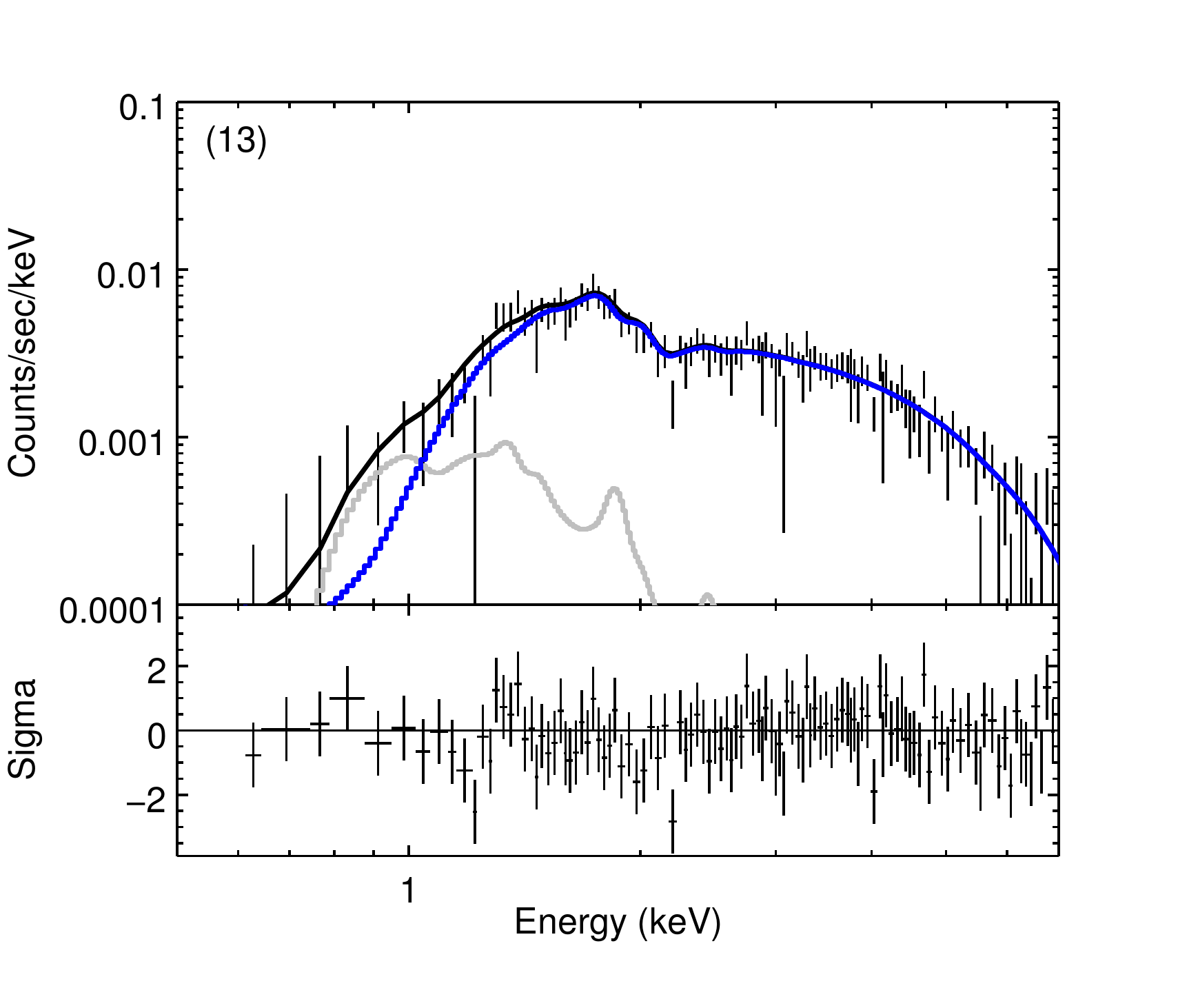}
\hspace{-6.5mm}
\epsscale{0.4} \plotone{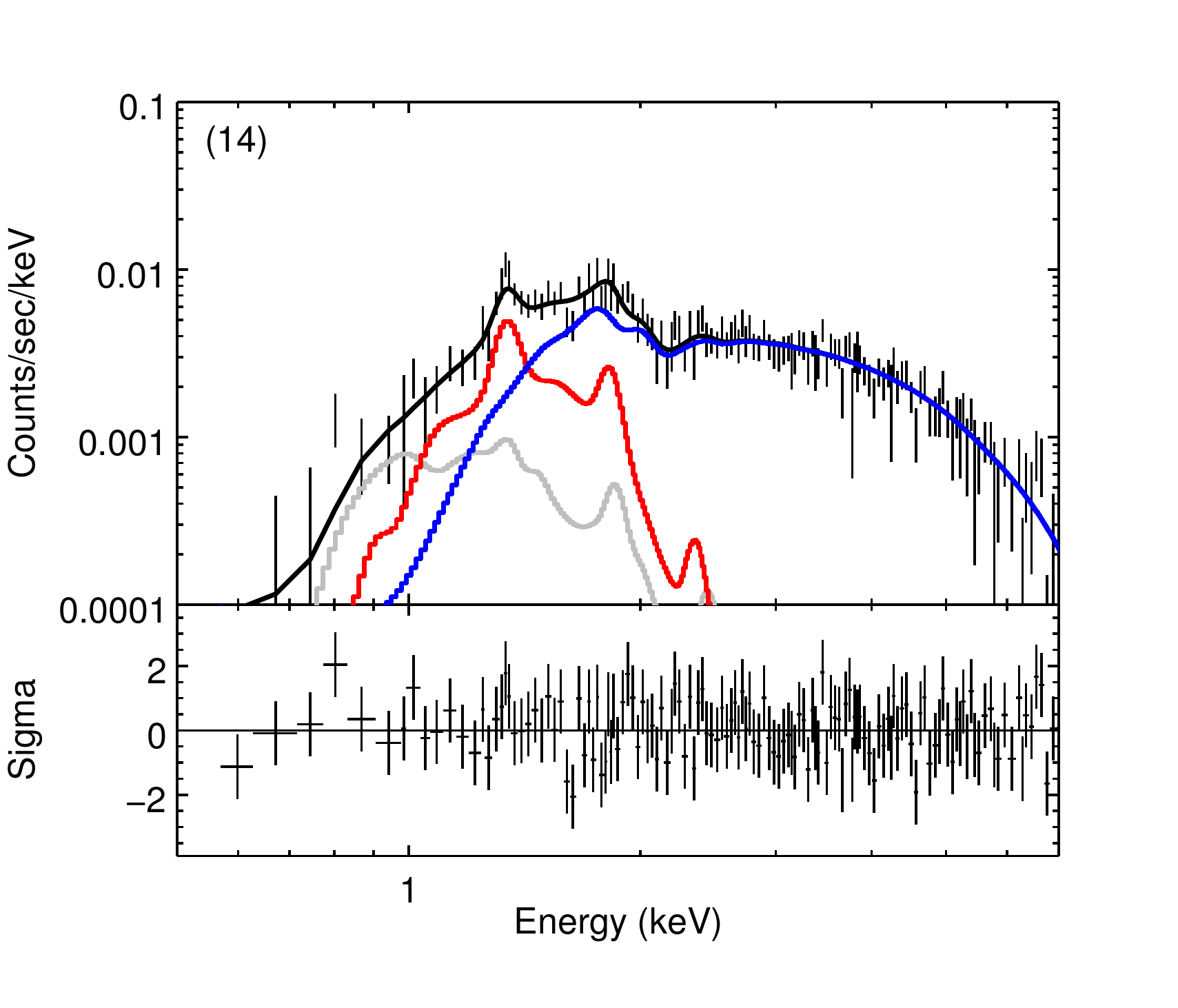}
\hspace{-6.5mm}
\epsscale{0.405} \plotone{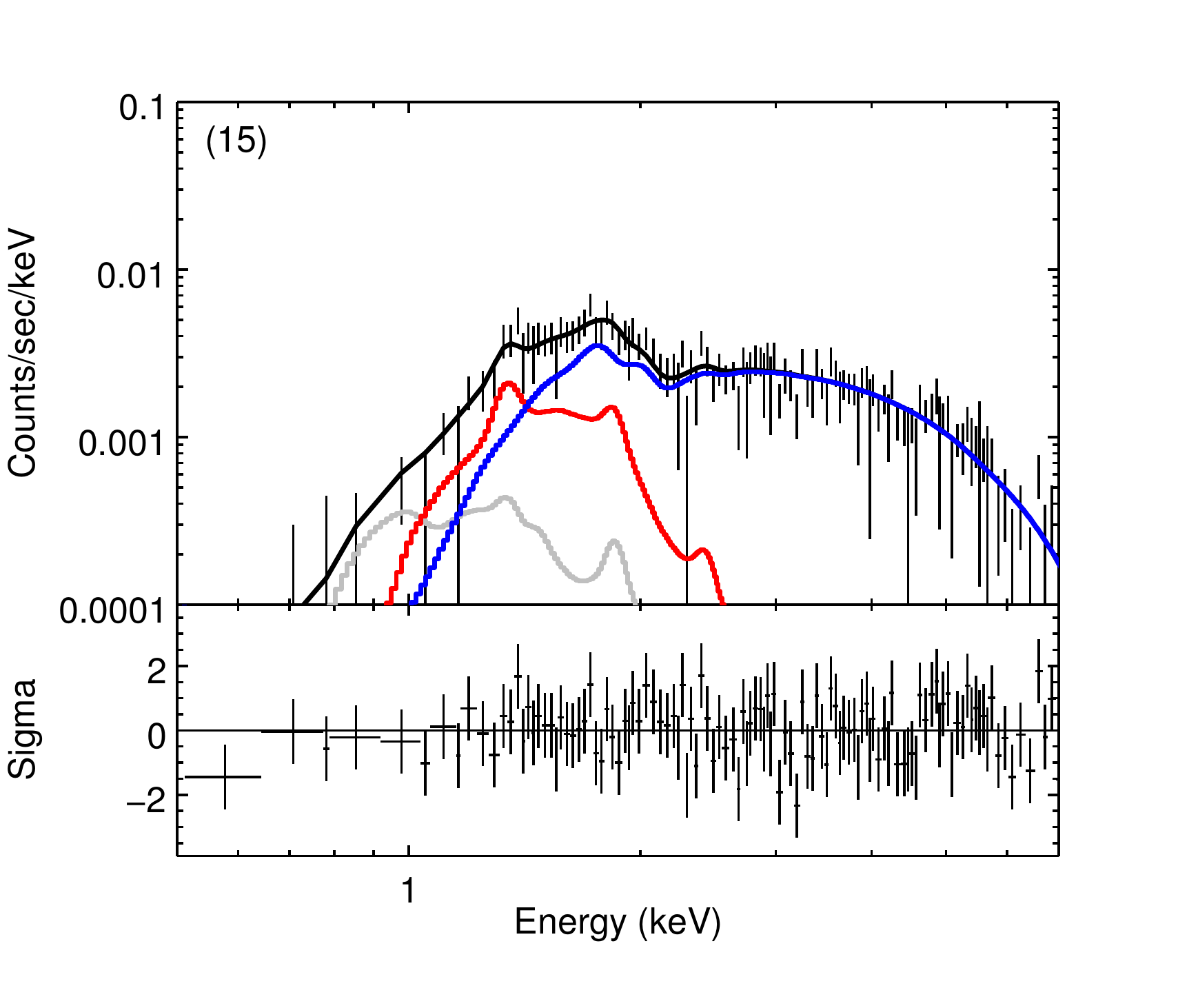}
\vspace{-4mm}
\epsscale{0.4} \plotone{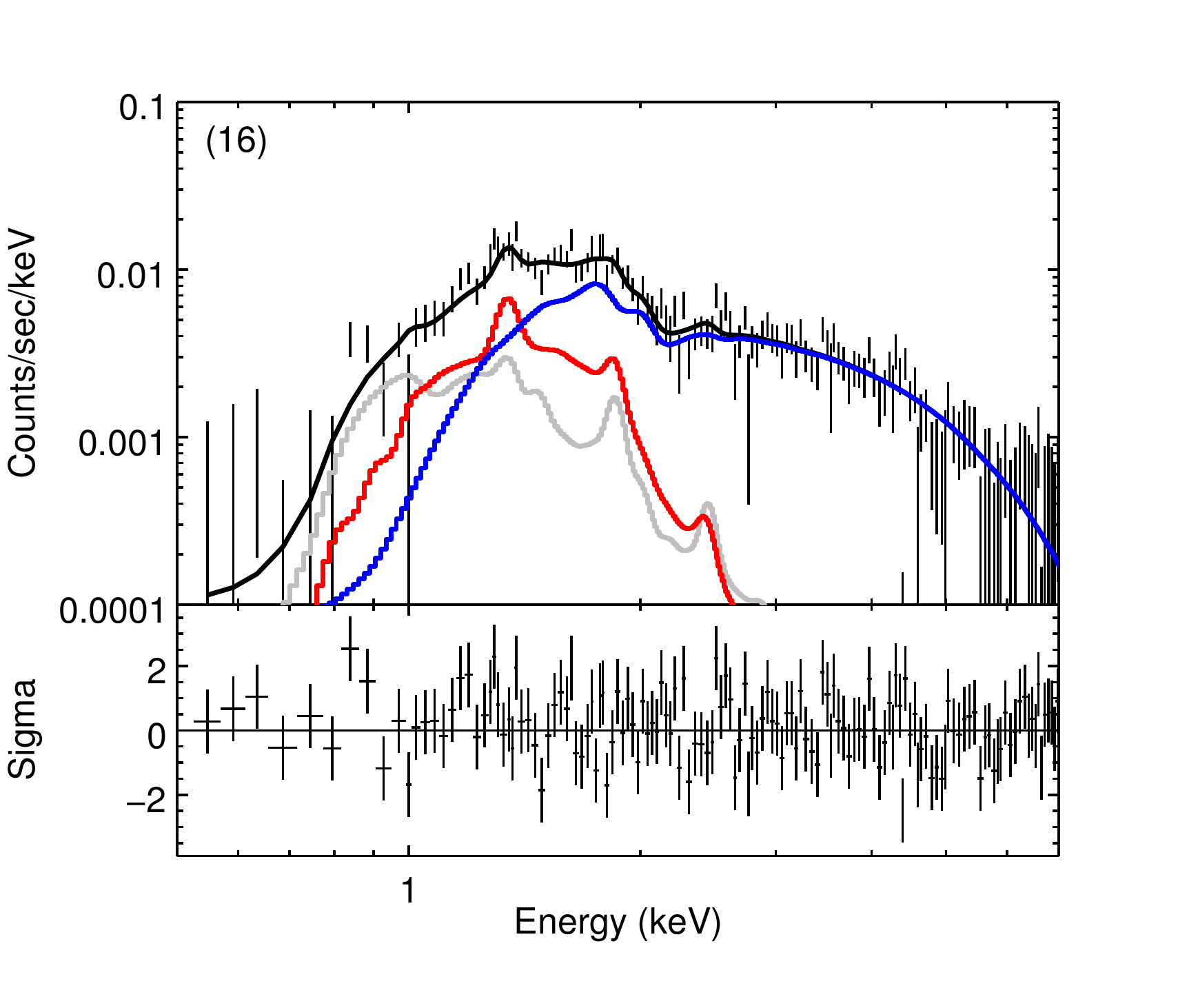}
\hspace{-6.5mm}
\epsscale{0.4} \plotone{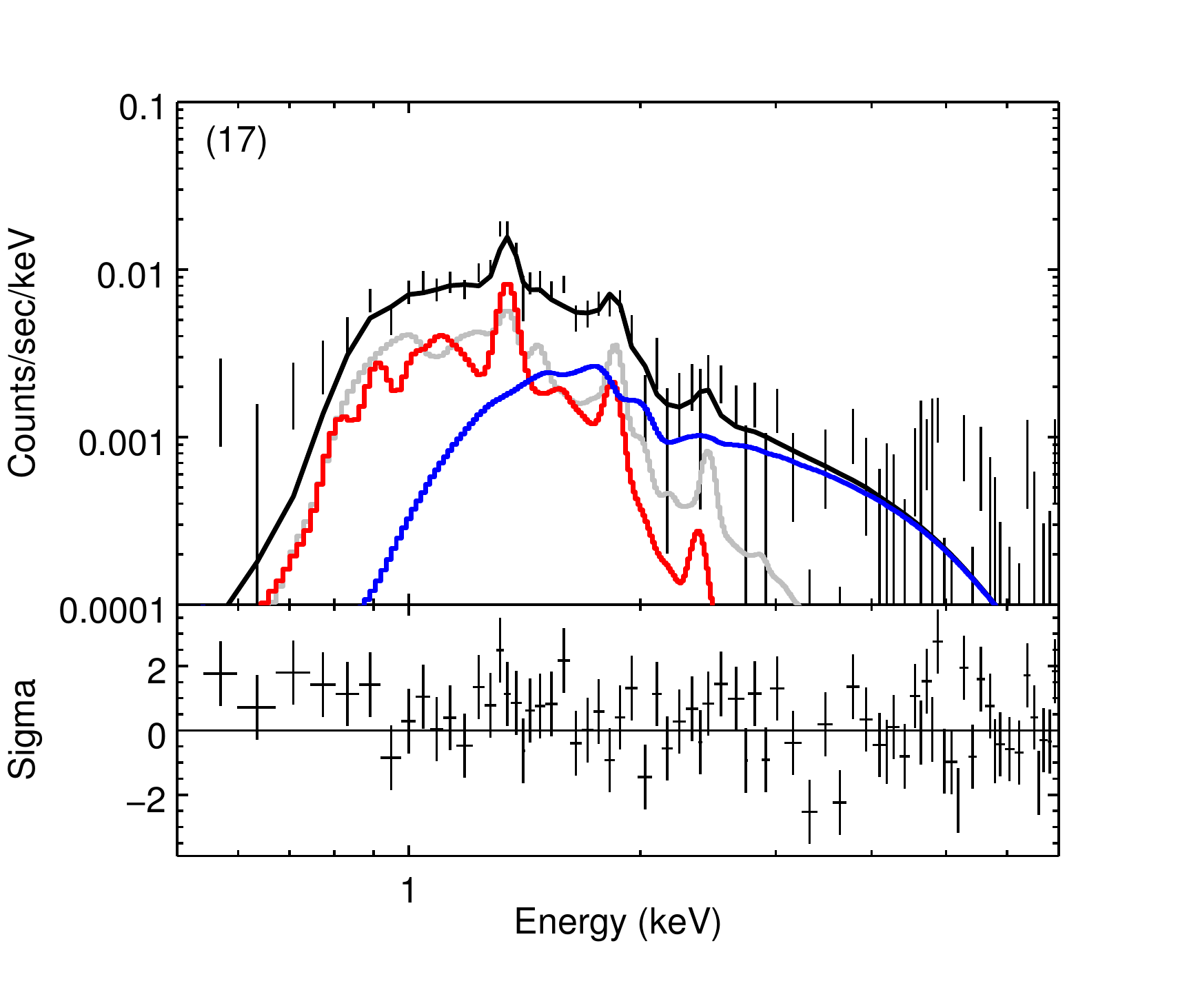}
\hspace{-6.5mm}
\epsscale{0.4} \plotone{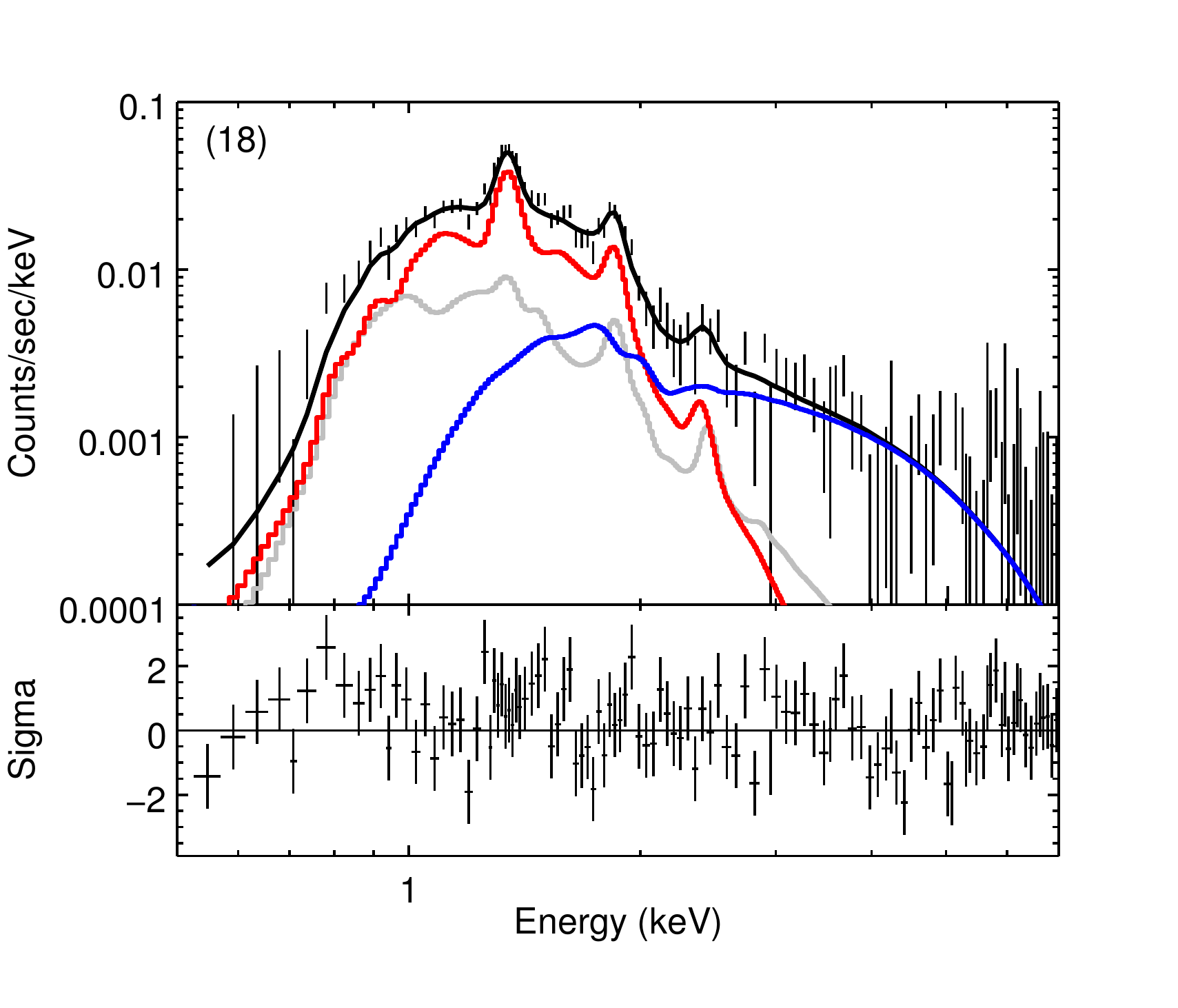}
\vspace{-4mm}
\epsscale{0.4} \plotone{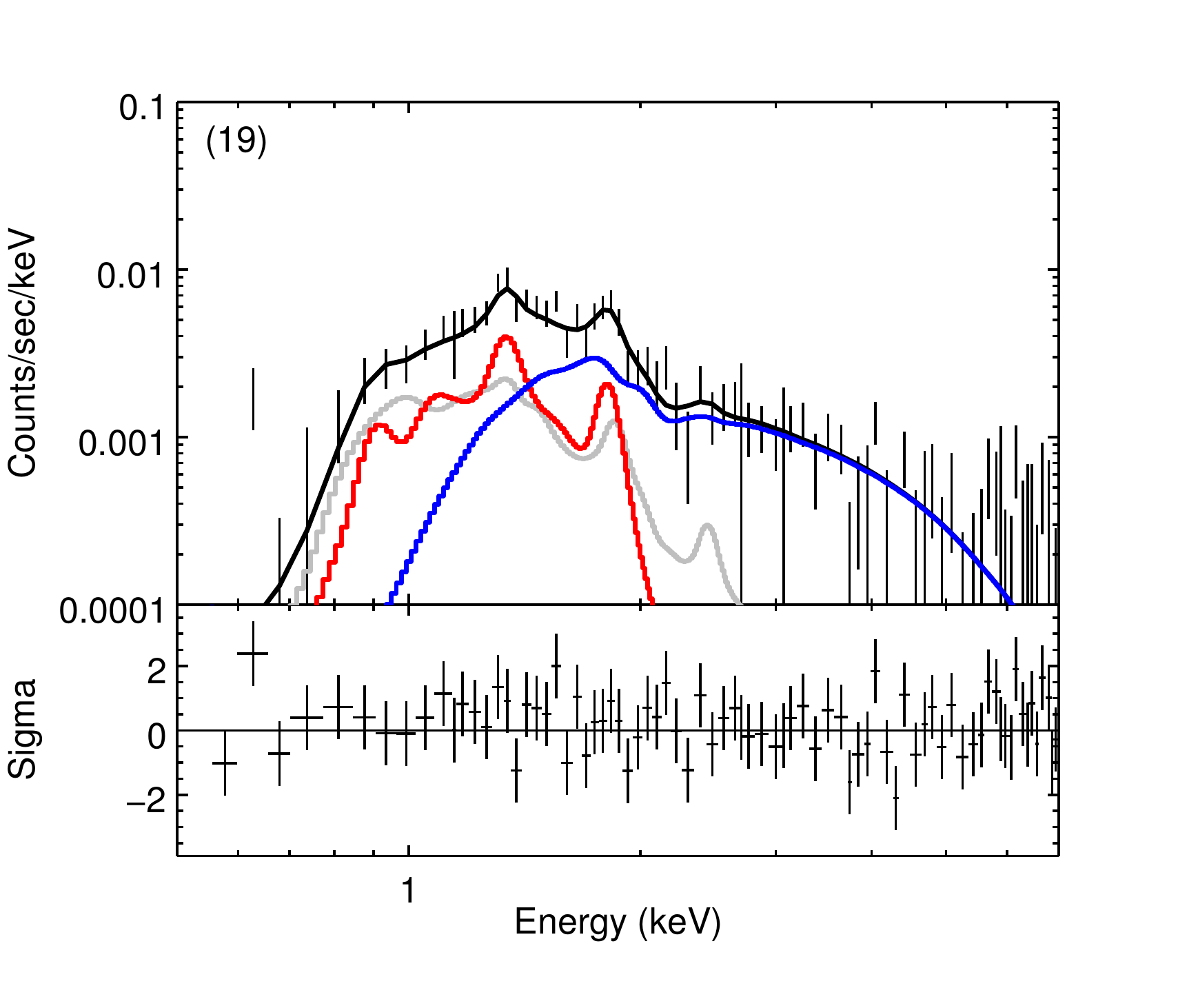}
\hspace{-6.5mm}
\epsscale{0.4} \plotone{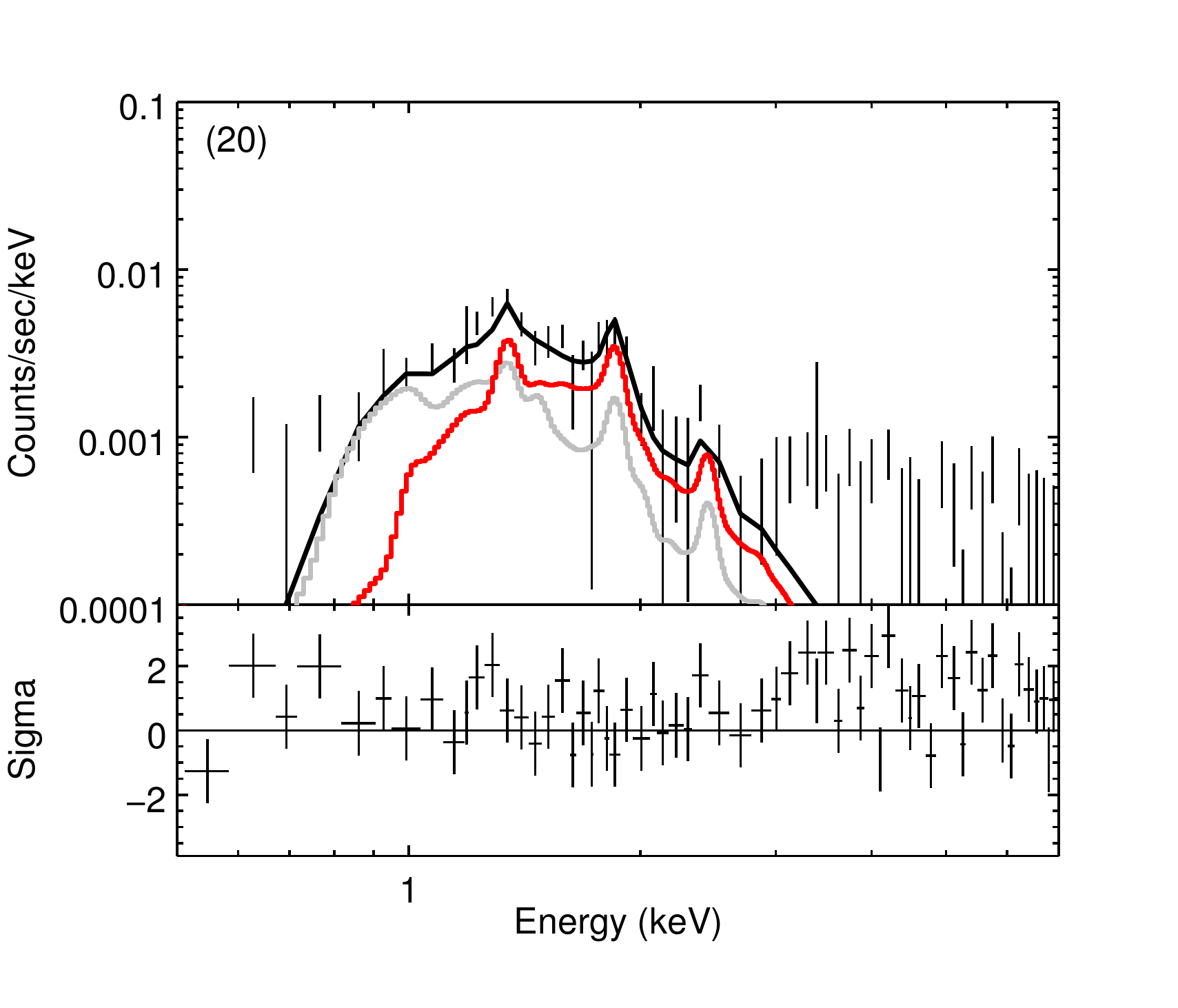}
\hspace{-6.5mm}
\epsscale{0.4} \plotone{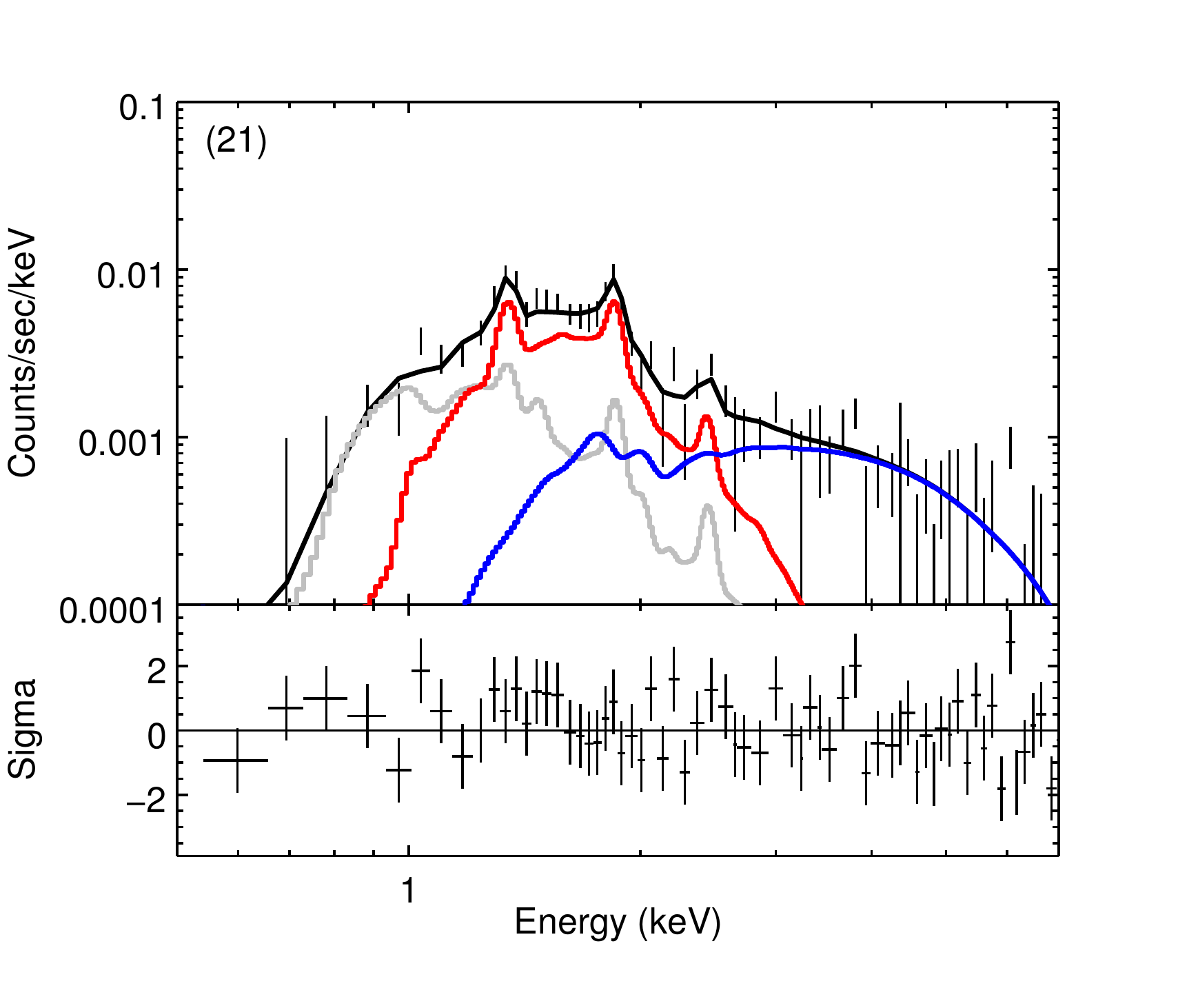}
\vspace{-4mm}
\epsscale{0.4} \plotone{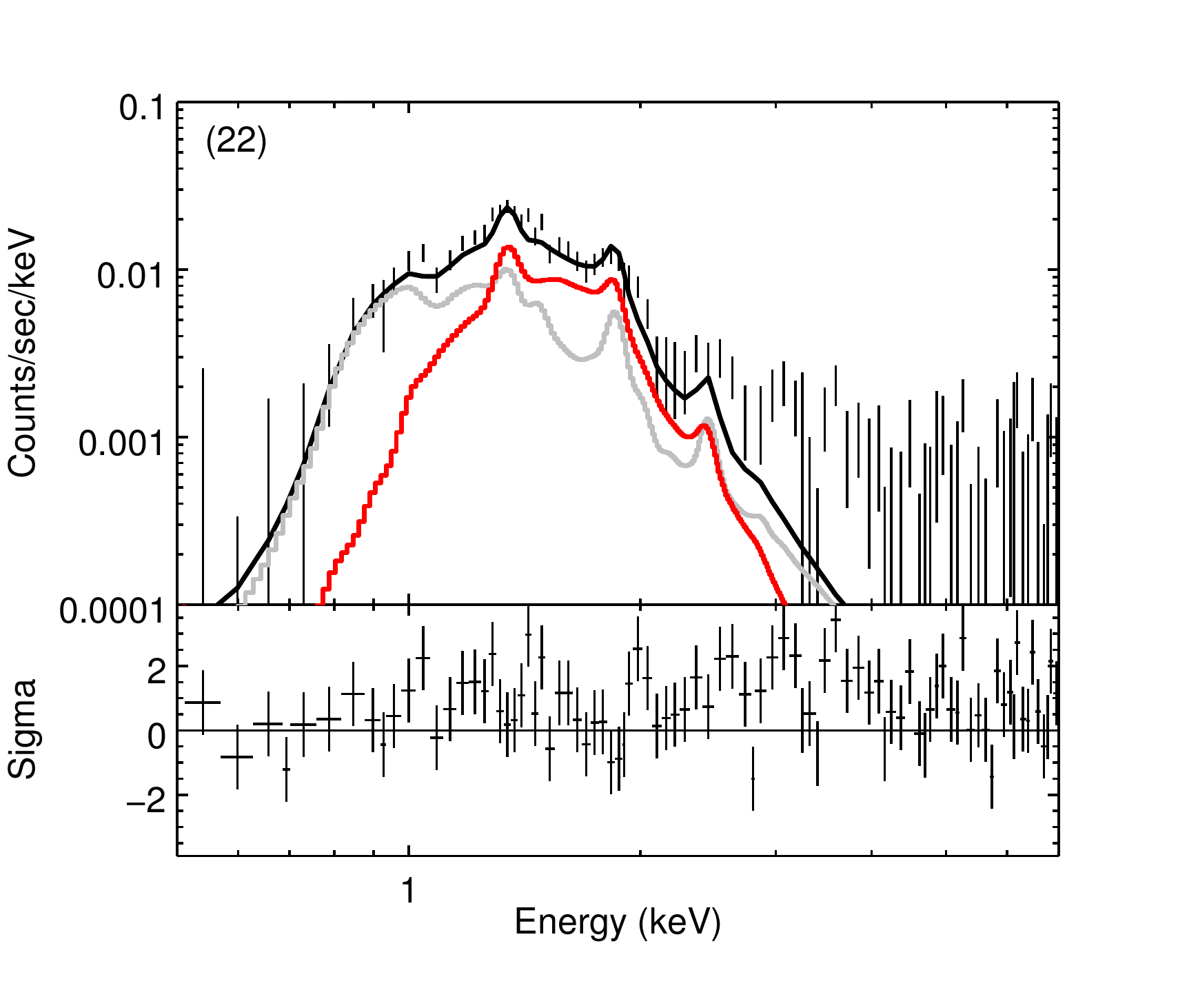}
\hspace{-6.5mm}
\epsscale{0.4} \plotone{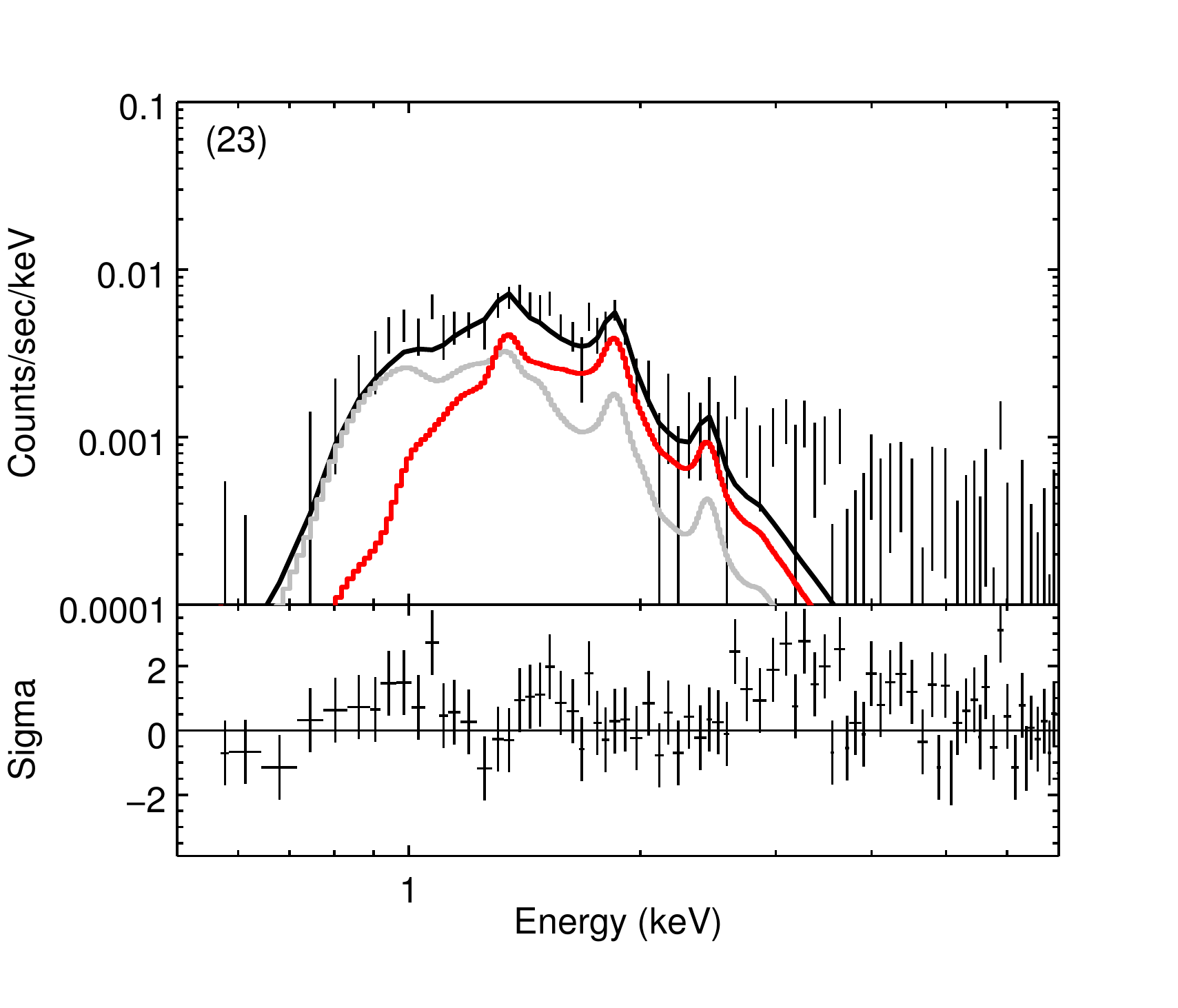}
\hspace{-6.5mm}
\epsscale{0.4} \plotone{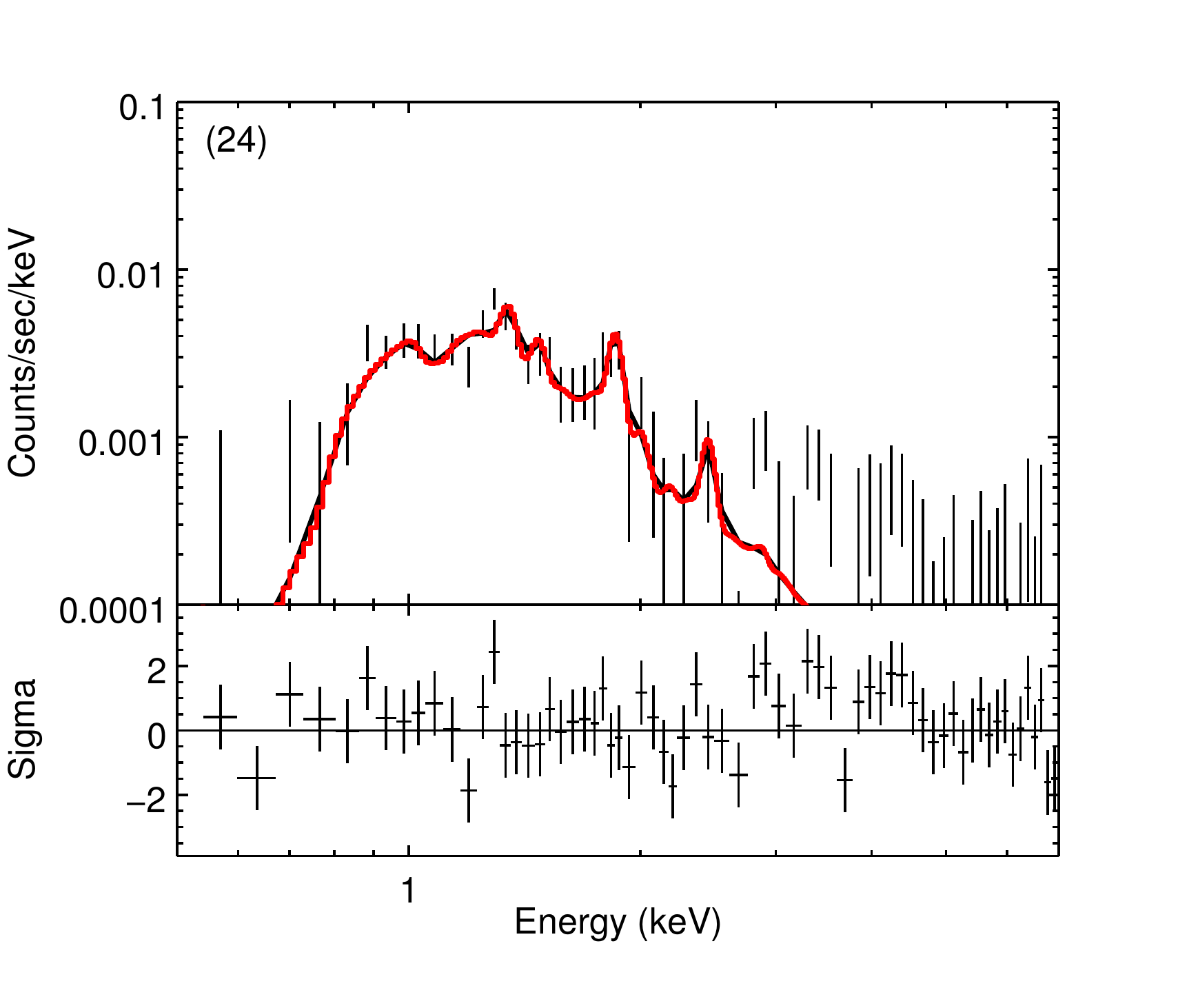}

\caption{\label{specfig3}Best-fit models and residuals the spectra from regions 13--24 shown in Figure~\ref{3color} and listed in Table~\ref{spec2}, where the numbers in the upper lefthand corners of the panels correspond to the region numbers. For all panels, the power-law component is shown in blue, thermal component in red, and residual background after blank-sky subtraction in gray (see Section~\ref{analysis}).}

\end{figure*}

In order to fit the source spectra from each extraction region, we first subtracted the corresponding blank-sky spectrum, and then included a model component for this additional thermal background excess, normalized by the extraction area for each region. The source and background spectra were then fitted simultaneously. All regions were fitted by either a power-law model describing the synchrotron emission from the PWN, a non-equilibrium ionization thermal model (\rm{XSVNEI}) with solar abundances \citet{anders89}, or both components for regions where both non-thermal and thermal emission was detected. The results of the fits are listed in Tables \ref{spec1} and \ref{spec2}. Table \ref{spec1} lists the best-fit parameters for the spectra extracted from the whole PWN and SNR shell, shown in Figure \ref{specfig1}. These regions are the white dotted regions in Figure \ref{3color}, where the region for the SNR shell is the outer circle with the inner PWN region excluded. Table \ref{spec2} lists the best-fit parameters for the numbered regions of Figure \ref{3color}. The spectra and best-fit models for all the regions are shown in Figures~\ref{specfig2} and \ref{specfig3}.

%**************************************************
%**************************************************

\section{General Morphology}\label{morph}

The radio morphology of G327.1-1.1 is that of a classic composite SNR; a faint shell $\sim$ 17\arcmin\ in diameter surrounds a bright PWN, approximately 5\arcmin\ in diameter, located off-center from the shell \citep{whiteoak96}. The radio contours are overlaid on the \textit{Chandra} X-ray Image in Figure~\ref{3color} (left). A distinct finger-like structure protrudes from the PWN in the northwest direction, possibly suggesting a picture in which a fast-moving pulsar is moving through the SNR, leaving the PWN in its wake. X-ray studies have confirmed the composite nature of G327.1-1.1, and provided evidence of a compact core that is the likely counterpart to the pulsar that powers the observed PWN emission \citep{lamb81,seward96,sun99,bocchino03,temim09}. Imaging in the hard X-ray band shows that the high energy photons are clearly offset from the main radio PWN in the direction of the radio finger, suggestive of a scenario in which the unusual morphology is produced by a combination of the pulsar's motion and the disruption of the PWN by the passage of the RS. \textit{XMM-Newton} observations of G327.1-1.1 showed a thermal X-ray shell, coincident with the radio shell emission, with brightest emission in the southeast \citep{temim09}. 

The new 350 ks \chandra observation presented in this work also shows soft X-ray emission that coincides with the radio shell, as shown by the red color in Figure~\ref{3color}. This diffuse emission is enhanced in the southeastern part of the shell, corresponding to region 18. The X-ray PWN is clearly displaced from the geometric center of the shell. The deep \chandra images show that the non-thermal emission from the PWN extends beyond the previously identified trail that coincides with the radio finger \citep{temim09}. The full extent of the hard X-ray emission from the PWN, shown in blue in Figure~\ref{3color}, approximately fills an ellipse with major and minor axis length of approximately 7\farcm9 and 4\farcm4, respectively. X-ray emission is now seen to completely fill the radio relic PWN, outlined in the contours of Figure~\ref{3color}. In addition to hard X-ray emission, the inner region of the radio relic also contains softer X-ray emission that is likely associated with thermal emission. While this emission may arise from the swept-up ISM seen in projection, its spatial correlation with the relic nebula suggests that it may originate from the inner SN ejecta that have mixed with the PWN material following the passage of the RS, as previously observed in systems such as Vela X \citep{lamassa08}.

As revealed by the earlier \chandra data, the brightest part of the PWN is the compact core at the tip of the radio finger which appears to contain the neutron star that powers the central PWN \citep{temim09}. The core is embedded in a bow-shock shaped cometary structure, with an extended X-ray trail connecting to the relic nebula, and prong-like structures that extend into large arcs ($\sim$ 2\farcm5 in length) towards the northwest (regions 13 and 14 in Figure~\ref{3color}) that appeared to form a bubble \citep{temim09}. The deep \chandra observation provides no evidence that the arcs actually connect to form a bubble, but instead suggests that this ``outflow" is an open structure.

\begin{deluxetable}{lccc}
\tablecolumns{4} \tablewidth{0pc} \tablecaption{\label{spec1}SPECTRAL FITTING RESULTS 1} 
\tablehead{ \colhead{PARAMETER} & \multicolumn{2}{c}{PWN} & \colhead{Shell}} 
\startdata

Area ($\rm arcsec^2$) & \multicolumn{2}{c}{207509} & 516889 \\

Counts & \multicolumn{2}{c}{147448} & 65420  \\

$\rm N_H$ ($\rm 10^{22}\:cm^{-2}$) & $1.76_{-0.07}^{+0.11}$ & $2.39_{-0.06}^{+0.21}$ &  $2.22_{-0.28}^{+0.14}$ \\

Photon Index & $2.15_{-0.04}^{+0.05}$ &  $2.27_{-0.02}^{+0.13}$ & $2.73_{-0.25}^{+0.23}$ \\

Amplitude ($10^{-3}$) & $3.58_{-0.19}^{+0.31}$ & $4.46_{-0.14}^{+0.78}$ & $1.45_{-0.41}^{+0.52}$ \\

$\rm F_1$ ($\rm 10^{-11}erg\:cm^{-2}\:s^{-1}$) & $1.88$ & $2.24$ & $0.71$ \\

kT (keV) & $0.29_{-0.09}^{+0.14}$ & $0.29_{-0.06}^{+0.01}$ & $0.29_{-0.04}^{+0.05}$ \\

$\rm [Mg]/[Mg]_\odot$ & $3.8_{-0.8}^{+2.5}$ & 1.0 &  $1.3_{-0.2}^{+0.5}$ \\

$\rm [Si]/[Si]_\odot$ & $5.1_{-3.3}^{+5.7}$ & 1.0 & $1.3_{-0.4}^{+1.1}$ \\

$\rm \tau$ ($\rm 10^{10}\:s\:cm^{-3}$) & $6.8_{-5.0}^{+12.8}$ & $0.57_{-0.24}^{+0.36}$ & $0.33_{-0.14}^{+...}$ \\

Norm. ($10^{-2}$) & $0.83_{-0.58}^{+3.09}$ & $6.51_{-0.87}^{+9.37}$ & $8.46_{-5.45}^{+14.2}$ \\

$\rm F_2$ ($\rm 10^{-11}erg\:cm^{2}\:s^{-1}$) & $4.8$ & $16$ & $24$ \\

Reduced $\chi^2$ & 0.9 & 1.0 & 1.0

\enddata
\tablecomments{The best fit parameters for a power-law plus non-equilibrium ionization thermal model (\rm{XSVNEI}). The listed uncertainties are 1.6 sigma (90 \% confidence) statistical uncertainties only. The fluxes were calculated in the 0.3-10.0 keV range. The amplitude for the power-law component is in the units of $\rm counts\: keV^{-1}\:s^{-1}$ (at 1 keV), and the normalization of the thermal model is equal to $10^{-14}n_e n_H V / 4\pi d^2 \: \rm cm^{-5}$, where $V$ is the volume of the emitting region and $d$ is the distance to the SNR. The parameter $\rm \tau$ represents the ionization timescale for the best-fit thermal model, and the number of counts for each region represents blank-sky-subtracted source counts. The fluxes $\rm F_1$ and $\rm F_2$ are in the units of $\rm erg\:cm^{-2}\:s^{-1}$ and they represent unabsorbed fluxes for the power-law and thermal components in the 0.3-10 keV band, respectively.}
\end{deluxetable}

\subsection{Structure of the Inner PWN}\label{pwnmorph}

The 350 ks \chandra observation presented in this work provides a much more detailed view into the structure of the inner X-ray PWN, contained in region 11 in the right panel of Figure~\ref{3color}. The zoomed in image of this inner region is shown in Figure~\ref{pwn}. A bright compact source, $\sim$~10\arcsec\ in diameter, is embedded in a 40\arcsec\ long cometary structure whose morphology has been suggestive to resemble a bow shock produced by a pulsar moving supersonically through the SNR interior \citep{temim09}. Behind the cometary region is a trail of X-ray emission, approximately 1\farcm4 in length, that spatially coincides with the radio finger and connects the cometary structure back to the radio nebula. We refer to this feature as the X-ray ``trail". A pair of narrow, well defined prong-like structures, $\sim$ 1\farcm8 in length, are seen on each side of the cometary structure, parallel with its long axis. These prong structures extend into the more diffuse arcs seen in Figure~\ref{3color}.

%%**************************************************

%**************************************************
%**************************************************

%**************************************************
%**************************************************

\subsubsection{The Compact Source}\label{compact}

Previous spatial modeling of the compact source itself showed that the emission can be described by a point source embedded in a more extended, slightly elliptical Gaussian component \citep{temim09}. When these best-fit components were subtracted from the original 50~ks \chandra image, there was an indication of an asymmetric residual that might suggest a more complex structure.
The new \chandra observation provides $\sim$~11,000 counts from the compact source, approximately seven times more than in the original observation. Approximately 5,000 of these counts are likely from the underlying cometary structure in which the compact source is embedded, and $\sim$~6,000 from the source itself. A spatial model of a point source embedded in an extended component leads to parameters consistent with the original results of \citet{temim09}. The point source contains $\sim$~580 counts, or $\sim$~10\% of the total compact source flux. It is embedded in an
extended component well described by a 2D Gaussian with a full width half maximum (FWHM) of $5\farcs7\pm0\farcs1$ and an ellipticity of $0.26\pm0.02$, containing $\sim$~5,400 counts. Even though the overall shape of this more extended component can be described by the elliptical Gaussian, its emission is not entirely smooth and symmetric, but instead shows evidence of a more complicated structure. The bottom three panels of Figure~\ref{pwn} show the unsmoothed zoomed-in image of the compact source on the left, the residual image of the extended component after the subtraction of a point source at the location of the black ``x" in the middle, and a deconvolved image of this same point-source-subtracted extended component on the right. The deconvolution was performed with the CIAO Sherpa task \textit{arestore} that uses the ACIS point spread function to recover spatial resolution. Both the original and deconvolved residual images (middle and right) clearly show structure within the compact source that contains the presumed pulsar, reminiscent of a torus structure observed around many other young neutron stars \citep{kargaltsev08}.

\section{Spectral Properties}\label{spectra}

The 50~ks \textit{Chandra} and 85~ks \textit{XMM-Newton} observations \citep{temim09} showed that the X-ray spectrum of the PWN can be described by a power-law with an average photon index of $2.11\pm0.03$ and the shell emission by a Raymond-Smith thermal plasma model with a temperature of 0.3~keV. There was also evidence for steepening of the photon-index with distance from the compact source that may be suggestive of synchrotron aging. Our deep \textit{Chandra} observations confirm the global spectral properties found in the previous work (see Table \ref{spec1}), but they allowed us to study the spectral properties of the individual PWN structures, and explore the spatial variations in the absorbing column density and the thermal emission from the SNR shell. Based on the best-fit parameters for the source spectra extracted from regions in Figure \ref{3color} and listed in Tables \ref{spec1} and \ref{spec2}, we make the following observations:

\begin{enumerate}[leftmargin=*]\itemsep0.0pt

\item The compact source and the cometary structure (regions 1 and 2) have photon indices that are consistent within the uncertainties, with $\Gamma=1.61^{+0.08}_{-0.07}$ and $\Gamma=1.62^{+0.08}_{-0.07}$, respectively. There is no evidence for significant steepening in the photon index in the cometary structure.

\item The photon index of the trail steepens from $\Gamma=1.76\pm0.12$ to $\Gamma=2.28\pm0.12$, giving a difference of $\Delta\Gamma(5-8)=0.52\pm0.17$, where the region numbers (Table \ref{spec2}) for which the difference in the photon index was calculated are indicated in the parentheses. As will be discussed in Section \ref{hdmodel2}, this can be explained by synchrotron aging of particles as they travel from the pulsar along the extent of the trail, a distance of approximately 80\arcsec\ (3.5 pc at a distance of 9 kpc). The trail may be interpreted as PWN material left behind by the pulsar as it travels through the SNR, along with being swept back by the passage of the SN reverse shock that effectively arrives at the pulsar from the NW direction, as a result of an east/west ISM density gradient (see Section \ref{hdmodel}).

\item The prong structures seem to extend southward of the compact source, parallel to the sides of the cometary structure. More specifically, regions 3 and 9 and regions 4 and 10 in Figure \ref{3color} seem to be part of the same structures (also see Figure \ref{pwn}). These structure have a photon index of $\sim$1.8. There is no evidence for spectral steepening along the prongs, with $\Delta\Gamma(3-9)=0.12\pm0.18$ and $\Delta\Gamma(4-10)=-0.05\pm0.18$.
The prongs extend into more diffuse and extended arc-like features (regions 13 and 14), approximately 155\arcsec\ in length from the northern tip of the prongs. Their synchrotron spectrum is significantly steeper, with a change in the photon index from the prong to the arc of $\Delta\Gamma(13-9)=0.58\pm0.28$ in the east and $\Delta\Gamma(14-10)=0.65\pm0.23$ in the west. The space between the arcs (regions 15 and 16) is also filled with synchrotron emission with a photon index similar to those of the arcs, suggesting that the whole structure may be a funnel shaped outflow, with the arcs possibly being limb brightened edges, rather than separate structures. The surface brightness of the PWN between the prongs (between regions 9 and 10) appears to be lower than in the surrounding nebula, which may be hinting that the funnel shaped outflow may be hollow.

%**************************************************

\begin{deluxetable*}{clccccccrrccc}
%\scriptsize
%\rotate
\footnotesize
\tablecolumns{12} \tablewidth{0pc} \tablecaption{\label{spec2}SPECTRAL FITTING RESULTS 2}
%\tablehead{
%\colhead{PARAMETER} & \colhead{Region 1} & \colhead{Region 2} & \colhead{Region 3} & \colhead{Region 4} & \colhead{Region 5} & \colhead{Region 6} & \colhead{Region 7} & \colhead{Region 8}}
%\startdata


\tablehead{
 \multicolumn{2}{c}{Region}  & \colhead{Area} & Cts & \colhead{$\rm N_H$}  & \colhead{Photon}  & \colhead{Amplitude}  &  \colhead{kT} & \colhead{$\rm \tau $} & \colhead{Norm.} & \colhead{$\rm F_{1}$} & \colhead{$\rm F_{2}$} & \colhead{Red.} \\
\colhead{} & \colhead{} & \colhead{($\rm arcsec^2$)} & (1000) & \colhead{($\rm 10^{22}\:cm^{-2}$)}  & \colhead{Index}  & \colhead{($10^{-4}$)}  & \colhead{$(\rm keV)$} & \colhead{$\rm (10^{12}\:s\:cm^{-3})$ } & \colhead{($\rm 10^{-3}$)} &  \multicolumn{2}{c}{($\rm 10^{-12}$)} & \colhead{$\chi^2$}
} 

\startdata
1 & Compact Source & 84.657 & 6.34 & $1.93^{+0.08}_{-0.08}$ & $1.61_{-0.07}^{+0.08}$ & $1.05_{-0.10}^{+0.11}$ & \nodata & \nodata & \nodata &$0.45$ & \nodata & 0.80 \\

2 & Cometary PWN  & 971.22 & 7.75 & 1.93 &  $1.62_{-0.07}^{+0.08}$ & $1.47_{-0.14}^{+0.16}$ & \nodata & \nodata & \nodata & $1.09$ & \nodata & \nodata \\

3 & Trail East  &  537.42 & 2.13 & 1.93  &  $1.84_{-0.12}^{+0.12}$ & $0.44_{-0.06}^{+0.07}$ &  \nodata & \nodata & \nodata & $0.27$ & \nodata & \nodata \\

4 & Trail West  & 766.56 & 3.12 & 1.93 &  $1.80_{-0.11}^{+0.11}$ & $0.61_{-0.08}^{+0.09}$ &  \nodata & \nodata & \nodata & $0.39$ & \nodata & \nodata \\

%5 & Trail Center  &  \nodata &  $1.97_{-0.10}^{+0.10}$ & $1.04_{-0.12}^{+0.13}\times 10^{-4}$ &  \nodata & \nodata & \nodata &  & \nodata & \nodata \\

5 & Trail 1  & 424.45 & 1.98 & 1.93 & $1.76_{-0.12}^{+0.12}$ &  $0.39_{-0.05}^{+0.05}$ & \nodata & \nodata & \nodata & $0.26$ & \nodata & \nodata \\

6 & Trail 2  & 588.19 & 2.13 & 1.93 & $1.95_{-0.11}^{+0.11}$ & $0.49_{-0.06}^{+0.07}$ & \nodata & \nodata & \nodata & $0.28$  & \nodata & \nodata \\

7 & Trail 3  & 994.92 & 2.99 & 1.93 & $2.09_{-0.10}^{+0.10}$ & $0.78_{-0.08}^{+0.09}$ & \nodata & \nodata & \nodata & $0.42$ & \nodata & \nodata \\

8 & Trail 4  & 839.48 & 2.38 & 1.93 & $2.28_{-0.12}^{+0.12}$ & $0.74_{-0.09}^{+0.09}$ & \nodata & \nodata & \nodata & $0.37$ & \nodata & \nodata \\

9 & Prong East  & 828.58 & 1.66 & 1.93 & $1.72_{-0.14}^{+0.14}$ & $0.30_{-0.05}^{+0.06}$ &  \nodata & \nodata & \nodata & $0.27$ & \nodata & \nodata \\

10 & Prong West & 971.22 & 2.06 & 1.93 & $1.85_{-0.14}^{+0.14}$ & $0.44_{-0.07}^{+0.08}$ & \nodata & \nodata & \nodata & $1.09$  & \nodata & \nodata \\

11 & Diffuse PWN$^{*}$  & 20007 & 27.7 & 1.93 & $2.11_{-0.05}^{+0.04}$ & $6.91_{-0.74}^{+0.37}$ & $0.23_{-0.05}^{+0.14}$ & $0.21_{-0.16}^{+0.88}$ & $6.0_{-4.0}^{+16}$ & $3.68$ & $17.7$ & 0.82 \\

12 & Relic PWN$^{*}$  & 26787 & 17.2 &  1.93 & $2.58_{-0.10}^{+0.07}$ & $6.51_{-0.71}^{+0.53}$ & 0.23 & $0.21$ &  $6.9_{-5.5}^{+18}$ & $3.14$ & $20.3$ & \nodata \\

13 & Arc East & 6147.6 & 4.60 & $2.08^{+0.23}_{-0.25}$ & $2.30_{-0.19}^{+0.20}$ & $1.62_{-0.38}^{+0.52}$ & \nodata & \nodata & \nodata & $0.81$ & \nodata & 0.74 \\

14 & Arc West & 6226.4 & 5.56 & $3.01^{+0.46}_{-0.44}$ & $2.50_{-0.19}^{+0.17}$ & $2.74_{-0.66}^{+0.74}$ & $0.21_{-0.04}^{+0.35}$ & $0.14_{-0.10}^{+0.22}$ & $34_{-23}^{+84}$ & $1.3$ & 102 & 0.95 \\

15 & Inner Arcs 1 & 3430.5 & 3.67 &  $2.94^{+0.71}_{-0.68}$ & $2.27_{-0.22}^{+0.27}$ & $1.61_{-0.45}^{+0.93}$ & $0.29_{-0.11}^{+0.52}$ & $0.69_{-0.54}^{+...}$ & $2.7_{...}^{+2.1}$ & $0.61$  & 6.0 & 0.88 \\

16 & Inner Arcs 2 & 22014 & 7.23 & $2.19^{+0.34}_{-0.41}$ & $2.52_{-0.20}^{+0.22}$ & $2.74_{-0.64}^{+1.09}$ & $0.29_{-0.11}^{+0.52}$ & $2.7_{...}^{+2.1}$ &  $3.1_{-2.6}^{+26}$ & $1.3$ & $7.2$ & 0.88 \\

%17 & NE Shell & 26361 & $2.47^{+0.16}_{-0.43}$ & \nodata & \nodata & $0.48_{-0.10}^{+0.1}$ & $40_{-...}^{+...}$ & $0.89_{-0.33}^{+0.11}$ & \nodata & $1.9$ & 1.00 \\

17 & East Shell & 42053 & 4.43 & $1.66^{+0.46}_{-0.28}$ & $2.85_{-0.92}^{+0.62}$ & $0.80_{-0.52}^{+0.64}$ & $0.30_{-0.09}^{+0.14}$ & $0.08_{-0.04}^{+0.24}$ & $2.2_{-1.7}^{+37}$ & 0.40 & 11.2 & 1.13 \\

18 & SE Shell & 62109 & 11.2 & $2.00^{+0.17}_{-0.08}$ & $2.73_{-0.54}^{+0.48}$ & $1.47_{-0.69}^{+1.61}$ & $0.31_{-0.07}^{+0.02}$ & $0.12_{-0.05}^{+0.06}$ & $13.3_{-3.6}^{+34.8}$ & 0.72 & $57$ & 1.02 \\

19 & North Shell & 18470 & 3.28 & $2.21^{+0.53}_{-0.70}$ & $3.01_{-0.69}^{+0.73}$ & $1.40_{-0.84}^{+1.94}$ & $0.18_{-0.07}^{+0.45}$ & $0.08_{-0.04}^{+0.4}$ & $49_{-48}^{+8800}$ & 0.73 & $130$ & 1.00 \\

20 & NW Shell & 23944 & 2.61 & $2.34^{+0.18}_{-0.50}$ & \nodata & \nodata & $0.53_{-0.14}^{+0.13}$ & $20.5_{-...}^{+...}$ & $0.85_{-0.73}^{+0.62}$ & \nodata & $1.5$ & 0.80 \\

21 & West Shell & 18227 & 3.40 & $3.00^{+0.23}_{-0.30}$ & $1.86_{-0.64}^{+0.62}$ & $0.34_{-0.12}^{+0.22}$ & $0.37_{-0.04}^{+0.05}$ & $16.4_{-16.3}^{+...}$ & $3.4_{-0.8}^{+2.0}$ & 0.21 & $6.9$ & 1.12 \\

22 & SW Shell & 69487 & 7.01 & $2.75^{+0.29}_{-0.14}$ & \nodata & \nodata & $0.29_{-0.05}^{+0.06}$ & $1.8_{-1.0}^{+...}$ & $11_{-6}^{+16}$ & \nodata & $22$ & 1.01 \\

23 & NE Shell & 26361 & 2.73 & $1.76^{+0.23}_{-0.32}$ & $2.46_{-0.42}^{+0.34}$ & $0.50_{-0.16}^{+0.35}$ & $0.37_{-0.08}^{+0.19}$ & $15_{-...}^{+...}$ & $0.36_{-0.24}^{+0.52}$ & 0.24 & $0.79$ & 0.80 \\

24 & Background & 19611 & 1.55 & $1.07^{+0.03}_{-0.09}$ & \nodata & \nodata & $0.68_{-0.05}^{+0.18}$ & $49_{-48}^{+...}$ & $0.08_{-0.02}^{+0.02}$ & \nodata & $0.21$ & 0.87 

%\sidehead{Fixed $\rm N_H = 1.91 \times 10^{22}cm^{-2}$}
%17 & Trail 1  &  \nodata & $1.77_{-0.12}^{+0.12}$ &  $3.92_{-0.51}^{+0.55}\times 10^{-5}$ & \nodata & \nodata & \nodata &  & \nodata & 0.52 \\
%18 & Trail 2  &  \nodata & $2.00_{-0.11}^{+0.11}$ & $5.03_{-0.61}^{+0.65}\times 10^{-5}$ & \nodata & \nodata & \nodata &  & \nodata & 0.64 \\
%19 & Trail 3  &  \nodata & $2.15_{-0.10}^{+0.10}$ & $8.21_{-0.85}^{+0.90}\times 10^{-5}$ & \nodata & \nodata & \nodata &  & \nodata & 0.70 \\
%20 & Trail 4  &  \nodata & $2.32_{-0.12}^{+0.12}$ & $7.52_{-0.88}^{+0.95}\times 10^{-5}$ & \nodata & \nodata & \nodata &  & \nodata & 0.69 \\
%21 & Prong East 1  &  \nodata & $1.80_{-0.19}^{+0.18}$ & $1.71_{-0.34}^{+0.38}\times 10^{-5}$ & \nodata & \nodata & \nodata &  & \nodata & 0.54 \\
%22 & Prong East 2  &  \nodata & $1.83_{-0.21}^{+0.21}$ & $1.63_{-0.36}^{+0.40}\times 10^{-5}$ & \nodata & \nodata & \nodata &  & \nodata & 0.56 \\
%23 & Prong West 1  & \nodata  & $1.95_{-0.18}^{+0.18}$ & $2.64_{-0.47}^{+0.56}\times 10^{-5}$ & \nodata & \nodata & \nodata &  & \nodata & 0.42 \\
%24 & Prong West 2  & \nodata & $1.97_{-0.19}^{+0.19}$ & $2.39_{-0.46}^{+0.52}\times 10^{-5}$ & \nodata & \nodata & \nodata &  & \nodata & 0.53 \\
\enddata
\tablecomments{Best-fit parameters for the spectra extracted from regions shown in Figure \ref{3color}, using a power-law and/or a non-equilibrium ionization thermal model (\rm{XSVNEI}). Listed uncertainties are 1.6 $\sigma$ (90 \% confidence) statistical uncertainties from the fit. The $\rm N_H$ value listed for Region 1 is the result of the simultaneous fit for regions 1--12. The parameters for the thermal components for Regions 11 and 12 were fitted simultaneously, as well as the thermal components for Regions 15 and 16. The parameters for the background region 24 is a simultaneous fit to the two background regions labeled ``24" in Figure \ref{3color}. The amplitude for the power-law component is in the units of $\rm counts\: keV^{-1}\:s^{-1}$ (at 1 keV), and the normalization of the thermal model is equal to $10^{-14}n_e n_H V / 4\pi d^2 \: \rm cm^{-5}$, where $V$ is the volume of the emitting region and $d$ is the distance to the SNR. The fluxes $\rm F_1$ and $\rm F_2$ are in the units of $\rm erg\:cm^{-2}\:s^{-1}$ and they represent unabsorbed fluxes for the power-law and thermal components in the 0.3-10 keV band, respectively. {*}The thermal component for the diffuse and relic PWN regions show a $\rm [Mg]/[Mg]_\odot$ overabundance of $2.31_{-0.61}^{+1.1}$. The thermal component in Region 24 is the excess background after blank-sky background subtraction. The number of counts for each region represents blank-sky-subtracted source counts.}
\end{deluxetable*}

%**************************************************

\item The steepest synchrotron spectrum in the PWN is observed in region 12, corresponding to the relic PWN observed at radio wavelengths. The best-fit photon index in this region is $\Gamma=2.6\pm0.1$. 
Fainter non-thermal emission also seems to be present in regions extracted from the SNR shell, away from the central PWN (regions 17, 18, 19, 21, 23). 
This non-thermal emission in the outer regions of the SNR is faint, at the level of only $\sim$ 2\% of the total PWN emission. It also has a photon index that is considerably steeper, leading us to conclude that it is most likely a dust-scattering halo from the PWN that is expected to have a softer spectrum due to the energy dependence of the scattering efficiency \citep[e.g.][]{bocchino05}.

\item The thermal emission both around the PWN and in the SNR shell can be characterized by the same average temperature of 0.29 keV (see Table \ref{spec1}). Within the uncertainties, there is no evidence that there are significant spatial variations in the thermal temperature across the SNR. The SNR properties derived using the \citet{sedov59} blast-wave model and the parameters from the best-fit thermal component are consistent with those found in \citet{temim09}; an ambient ISM density $n_0=0.12\:\rm cm^{-3}$, a shock velocity $v_s= 500 \rm \:km\:s^{-1}$, an explosion energy $E_{51}(10^{51}\rm erg\:s^{-1})=0.5$, and an age of 17400 yr. The estimates assume an SNR distance of 9.0 kpc.

The spectrum extracted from the entire PWN region is well fitted by both a model with fixed solar abundances, and a model with enhanced abundances of Mg and Si. While the enhanced abundance model provides a slightly better fit, the difference in reduced $\chi^2$ values is not significant enough to distinguish between the two models. Both spectral fits are shown in Figure \ref{specfig1} and parameters listed in Table \ref{spec1}.
The enhanced abundance model gives best fit values of $3.8^{+2.5}_{-0.8}$ and $5.1^{+5.7}_{-3.3}$ times solar, for Mg and Si respectively, and would suggest an interaction of the PWN with SN ejecta, or possible mixing of the SN ejecta and PWN material following a RS interaction. As mentioned in the previous section, the morphology of the soft X-ray emission inside the relic PWN, supports the idea that the soft thermal emission in this region originates from SN ejecta. This is also supported by the fact that the best-fit ionization timescale for the solar abundance model is implausibly low for emission arising from swept-up ISM material in an evolved SNR. The most likely scenario is that this region contains a mix of SN ejecta and ISM/CSM emission that is either projected emission from shocked SNR shell, ISM material that has mixed into the SNR interior, or both.

\item The average best-fit column density is $\rm N_H=1.76_{-0.07}^{+0.11}\times10^{22}\:cm^{-2}$ for the PWN region and $\rm N_H=2.22_{-0.28}^{+0.14}\times10^{22}\:cm^{-2}$ for the SNR shell (see Figure \ref{3color} and Table \ref{spec1}). There is an indication of a somewhat higher absorbing column density in the western side of the SNR shell (regions 14, 21, and 22). This higher $\rm N_H$ value along the line of sight in the west is not necessarily intrinsic to the SNR, but it is consistent with a scenario in which the SNR expands into denser ambient medium in the west.  

\end{enumerate}

%**********************

\begin{deluxetable}{llc}
\tablecolumns{3} \tablewidth{20pc} \tablecaption{\label{properties}SNR PROPERTIES \& HD MODEL INPUT PARAMETERS} 
\tablehead{\colhead{Parameter} & \colhead{Description} & \colhead{Value}} 
\startdata
\\
\underline{SNR Properties:} & & \\
$D$ (kpc){*} & SNR distance & 9.0 \\
$R_{SNR}$ (pc){*} & SNR radius & 22 \\
$v_s$ (km/s){*} & Shock velocity & 500 \\
$t$ (yr){*}  & SNR age &17400 \\
$n_0$ ($\rm cm^{-3}$) & Average ambient density & 0.12 \\
$E_{51}$ ($\rm 10^{51}\:erg$) & Explosion energy & 0.5 \\
$M_{ej}$ ($\rm M_{\odot}$) & SN ejecta mass & 4.5 \\
\\
\underline{PWN Properties:} & & \\
$R_{PWN}$ (pc){*} & PWN radius & 5.0 \\
$v_p$ (km/s) & Pulsar velocity & 400 (north) \\
$L_{X(2-10)}$ (erg/s) & PWN X-ray luminosity & $7.2\times10^{34}$ \\
$\dot{E_0}$ (erg/s) & Initial spin-down luminosity & $2.8\times10^{38}$ \\
$n$ & Pulsar braking index & 3.0 \\
$\tau_0$ (yr) & Spin-down timescale & 2000 \\
$B$ ($\rm \mu G$) & PWN magnetic field & 11 \\
\\
\underline{Density Gradient:} & & \\
$x$ & Density contrast of 12.5 & 1.08 \\
$H$ (pc) & Characteristic length scale &  5.2 \\
Orientation & & East/West \\
\enddata
\tablecomments{The SNR properties were derived using the Sedov model and the best-fit thermal parameters, assuming a distance of 9.0 kpc (Section~\ref{spectra}). The magnetic field $B$ was estimated from the model in Section~\ref{broadband}, while the pulsar velocity $v_p$ was estimated in \citet{temim09}. Parameters marked with an asterisk are not inputs to the HD simulation. The ejecta mass $M_{ej}$, pulsar braking index, initial spin-down luminosity $\dot{E_0}$, and the spin-down timescale $\tau_0$ were adjusted to produce the desired PWN morphology and SNR/PWN dimension at the derived SNR age of $\sim$ 17000 yr. The density gradient function is described by equation 14 of \citet{blondin01}.}
\end{deluxetable}

%**********************

%**************************************************
\begin{figure*}
\epsscale{0.9} \plotone{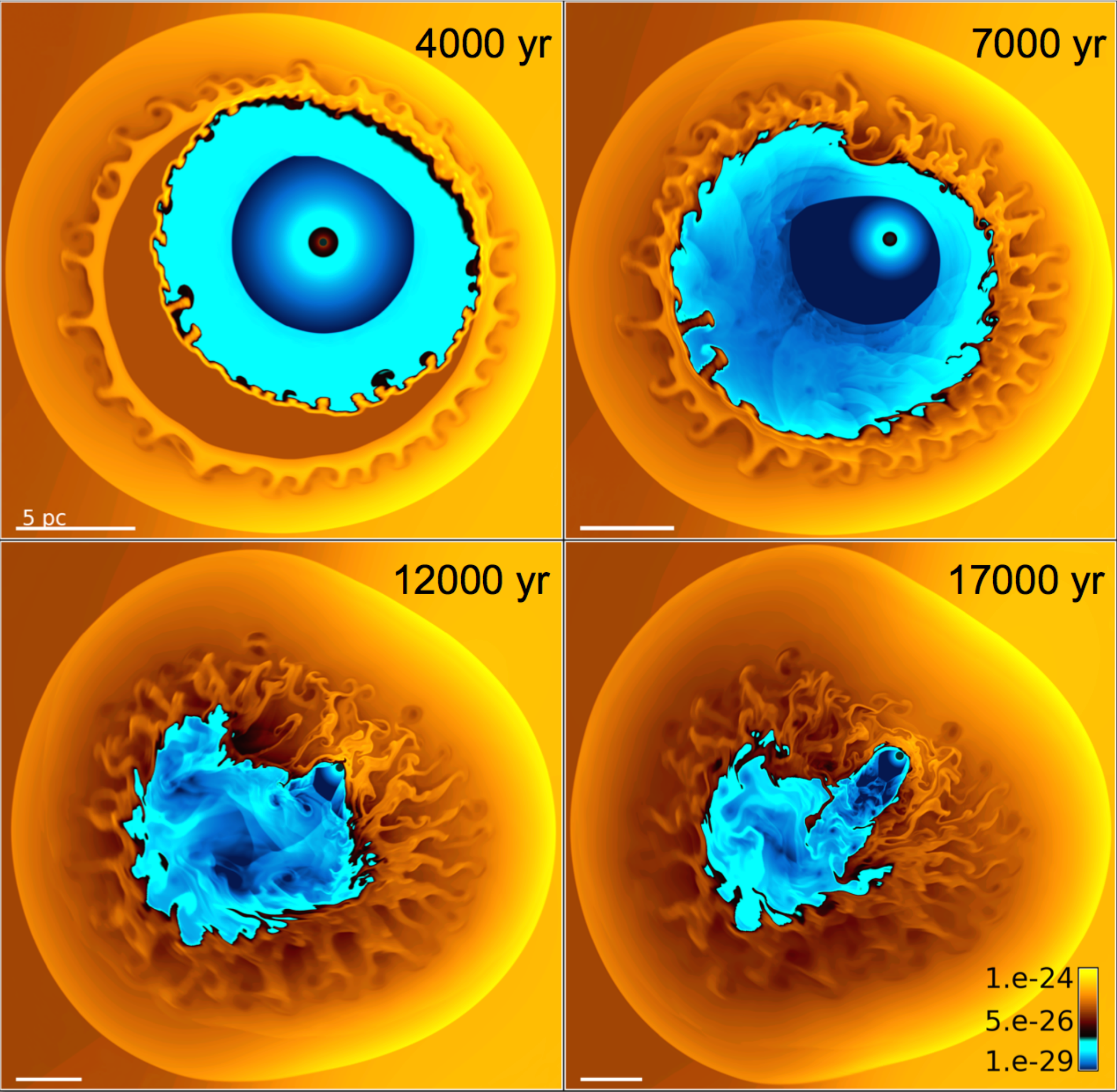}
\caption{\label{hdsim}Results of the HD simulation of a composite SNR expanding in an ambient density gradient, and containing a rapidly moving pulsar. The figure shows density maps ($\rm g\: cm^{-3}$) at four different SNR ages. The density is higher in the west and decreases towards the east. The pulsar is moving to the north at a velocity of 400 $\rm km\:s^{-1}$. The input parameters for the simulation are consistent with observed values for G327.1-1.1. The resulting morphology at 17000 years closely resembles the morphology of G327.1-1.1 at radio and X-ray wavelengths.}
\end{figure*}
%**************************************************

\section{Hydrodynamical Modeling}\label{hdmodel}

In order to understand the morphology of G327.1-1.1, we have modified the VH-1 hydrodynamics code\footnote{http://wonka.physics.ncsu.edu/pub/VH-1/} \citep{blondin01}
to model the evolution of a composite SNR that expands into a non-uniform ambient density and contains a rapidly moving pulsar.
The 2D simulation is evolved on a cylindrical grid (r,$\theta$), but with spherical radial divergence (volume proportional to $r^3$).
The model is constructed within the pulsar's rest frame and consists of a stationary, expanding gas bubble with an adiabatic index $\gamma=4/3$ for relativistic gas, contained within an SNR modeled by a self-similar driven wave expanding into an SN ejecta density profile described in \citet{blondin01}. Since the pulsar is stationary in the grid's reference frame, the SNR and the surrounding ISM were given a negative velocity to simulate the pulsar's true motion. This choice of frame is constructed to utilize the inner boundary to act as both an energy injection point for the pulsar wind and as a hard-surface modified to keep large amounts of mass from exiting the grid during the collision with the RS.

The model assumes a time-dependent pulsar spin-down power $\dot{E}= \dot{E_0}(1+t/\tau_0)^{-(n+1)/(n-1)}$ that drives the PWN, where $\tau_0$ is the pulsar's spin-down timescale and $n$ is the braking index. The magnetic field and braking index are assumed to be constant, as presented in \citep{blondin01}. The model also includes a gradient within the ambient ISM density that can be rotated about the grid center to achieve a desired angle between the orientation of the gradient and the pulsar's direction of motion.

\subsection{Input and Output Parameters}

The input parameters for the HD simulation are summarized in Table~\ref{properties}. We used an explosion energy of $0.5\times10^{51}\rm\:erg$ and an average ambient density of $0.12\rm\:cm^{-3}$, as estimated from the Sedov model and the best-fit thermal parameters for G327.1-1.1 (see Section \ref{spectra}). We assumed an ISM density gradient that decreases from west to east, as suggested by our best-fit measurements of the absorbing column density $N_H$. The pulsar is assumed to be moving to the north with a velocity of 400 $\rm km \: s^{-1}$ \citep{temim09}.

%We assumed an ISM density gradient described by equation 14 of \citet{blondin01}, with $x=1.08$, corresponding to a density contrast of 12.5, and the characteristic length scale over which the density changes ($H$) set to 5.2 pc. 
%%
%%We assumed a density gradient that decreases from west to east, as suggested by our best-fit measurements of the absorbing column density $N_H$. The pulsar is assumed to be moving to the north with a velocity of 400 $\rm km \: s^{-1}$ \citep{temim09}.

The parameters for which we did not have observational constraints were adjusted to produce a morphology that most closely resembles that of G327.1-1.1 at its estimated age of $\sim$~17000 yr. 
We assumed an ISM density gradient profile described by equation 14 of \citet{blondin01}, with $x=1.08$, corresponding to a density contrast of 12.5, and the characteristic length scale over which the density changes ($H$) set to 5.2 pc.
For the PWN, we fixed the braking index to $n=3.0$, and found an initial spin-down luminosity of $\dot{E}_0=2.8\times10^{38}\:\rm erg\:s^{-1}$ and an initial spin-down timescale of the pulsar $\tau_0=2000\rm \:yr$. This leads to a current spin-down luminosity $\dot{E}=3.1\times10^{36}\rm\:erg\:s^{-1}$. The observed non-thermal luminosity of the PWN in G327.1-1.1 is $L_X\rm(2-10\: keV) = 7.2\times10^{34}\rm\:erg\:s^{-1}$, for a distance of 9 kpc, which gives $\dot{E}/L_X=0.023$. This ratio is within the range observed for other pulsars \citep{possenti02}. The total SN ejecta mass used for the final run is 4.5 $\rm M_\odot$. The outputs of the simulation are the total pressure and density maps at each time step, as well as maps that separately trace the ISM, SN ejecta, and PWN material. We also trace the age of the particles injected by the pulsar in order to determine where in the PWN we should expect spectral steeping due to synchrotron aging. This can then be compared with the spectral steepening that we measure from the PWN spectra in Section \ref{spectra}.

\subsection{Comparison with G327.1-1.1}\label{hdmodel2}

The results of the HD simulation at four different ages are shown in Figure~\ref{hdsim}. The combined effect of the higher ambient density in the west and the northward displacement of the PWN due to the pulsar's motion result in an asymmetric interaction of the RS with the PWN from the NW direction. The interaction begins at an age of 3000 yr, with the RS reaching the entire surface of the PWN at around 7000 yr. At this stage, the Rayleigh-Taylor instabilities at the PWN/RS interface also become more prominent. The RS sweeps the PWN away from the pulsar and displaces it eastward of the explosion center. Once the RS reaches the pulsar itself, at an age of $\sim$~12000 yr, a narrow finger/trail of PWN material begins to form. Since the pulsar continually injects wind-producing particles as it moves to the north, in the absence of any density gradient, the PWN trail would be oriented northward. However, in our case, the northward motion of the pulsar combined with the preferential arrival of the RS from the west causes the trail to be oriented along the NW direction. 

The simulation suggests that the displacement of the relic PWN in G327.1-1.1 is mostly determined by the orientation of the ambient density gradient, while the orientation of the trail behind the pulsar is determined by the combination of the density gradient and pulsar's velocity. The formation and thickness of the trail also strongly depends on the explosion energy and the pulsar's spin-down luminosity. A spin-down time allowing for an initial spin-down luminosity below $10^{37}\rm\: erg\:s^{-1}$ will generally struggle to displace the incoming shock and form a narrow trail, while a higher energy output will easily maintain a larger bubble around the pulsar. None of the simulation runs produced non-thermal structures ahead of the pulsar that might explain the prongs and arcs observed in the \textit{Chandra} images.

At an age of 17000 yr, the Sedov estimated age for G327.1-1.1, the morphology closely resembles the radio and X-ray observations of the SNR. The physical sizes of the SNR and PWN radii correspond well to those measured for G327.1-1.1 assuming a distance of 9 kpc. The simulation also shows that the SN ejecta mixes with the PWN material during the RS interaction, as suggested by the observed morphology of the soft X-ray emission in Figure~\ref{3color}.

\begin{figure}
\center
\epsscale{0.8} \plotone{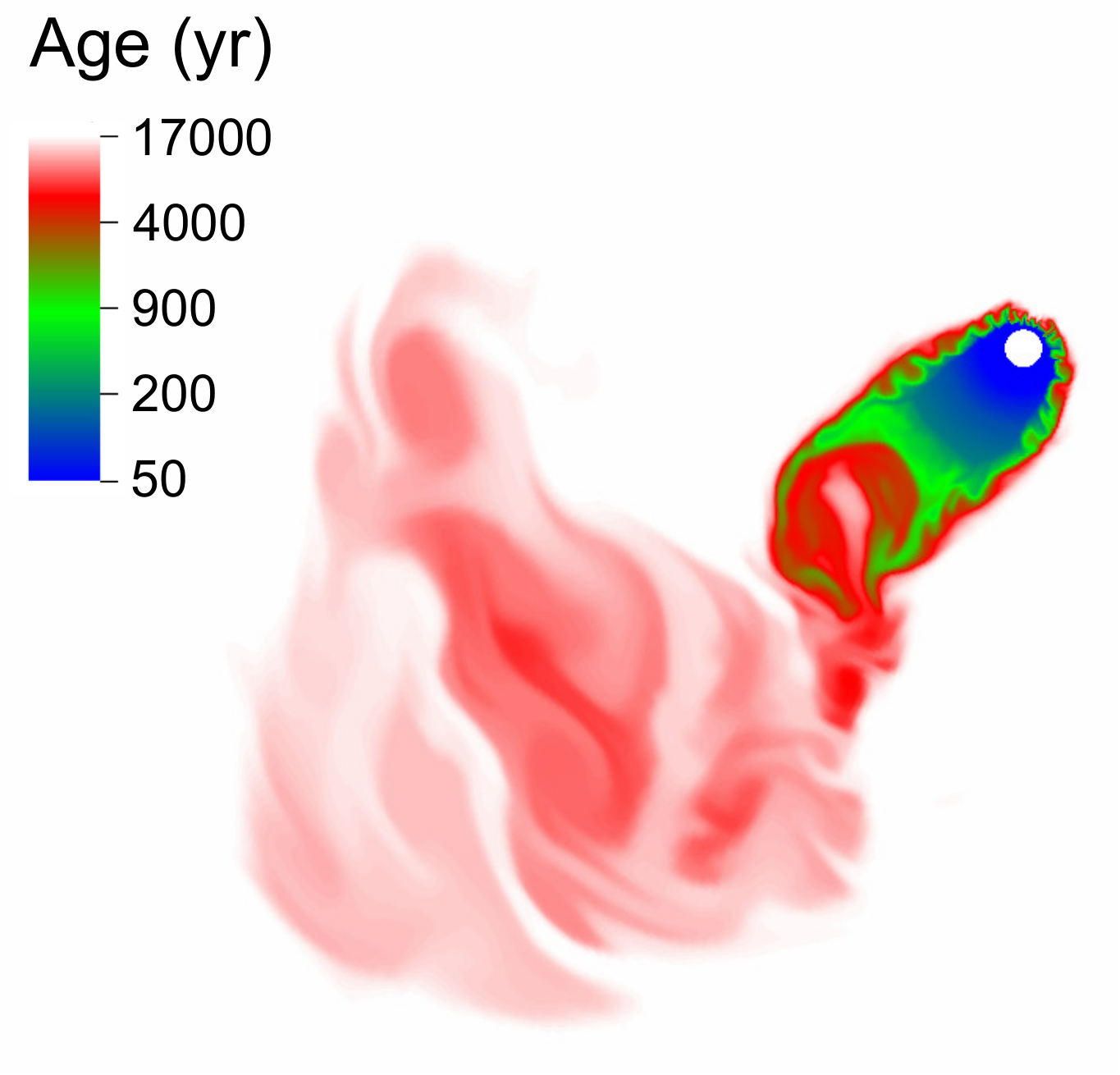}
\caption{\label{pwnage}Map of the PWN from the HD simulation described in Section \ref{hdmodel}, showing the age of the particles injected by the pulsar at an SNR of 17000 years. The region shown is the same as the blue colored PWN region of the lower right panel of Figure~\ref{hdsim}. The freshly-injected particles (blue) give rise to the X-ray emission in the cometary structure and the trail, while the aged PWN material shown in red corresponds to the morphology of G327.1-1.1 observed at radio wavelengths.}

\end{figure}

The simulation also keeps track of the age of the particles injected by the pulsar. Figure \ref{pwnage} shows the map of the particle age parameter at the present SNR age of 17000 yr. The freshly-injected PWN material is shown in blue, and the older particles in the light shades of red. As expected, the morphology of the aged particles in red resembles the radio emission from the PWN in G327.1-1.1, while the distribution of younger particles in blue and green traces the X-ray emission along the trail. 
Our analysis of the X-ray spectroscopy shows a steepening in the X-ray trail of $\Delta\Gamma=0.52\pm0.17$ (Section~\ref{spectra}), consistent with the steepening expected from synchrotron aging of particles as they travel away from the pulsar. Using a magnetic field of 11 $\rm \mu G$ (see Section~\ref{broadband}), we estimate that the synchrotron lifetime over which the steepening is expected to occur is $\tau_{syn}=1700\rm\: yr$. This means that in order to observe a spectral steeping of $\Delta\Gamma=0.5$ along the trail, the particles would have to had traveled for at least 1700 yr. Figure \ref{pwnage} shows that the particles at the base of the trail are $\sim$~4000 yr old, consistent with the scenario in which the observed steeping in the X-ray trail is caused by synchrotron aging.

%**************************************************
%**************************************************

%**************************************************
\begin{figure}
\center
\epsscale{1.1} \plotone{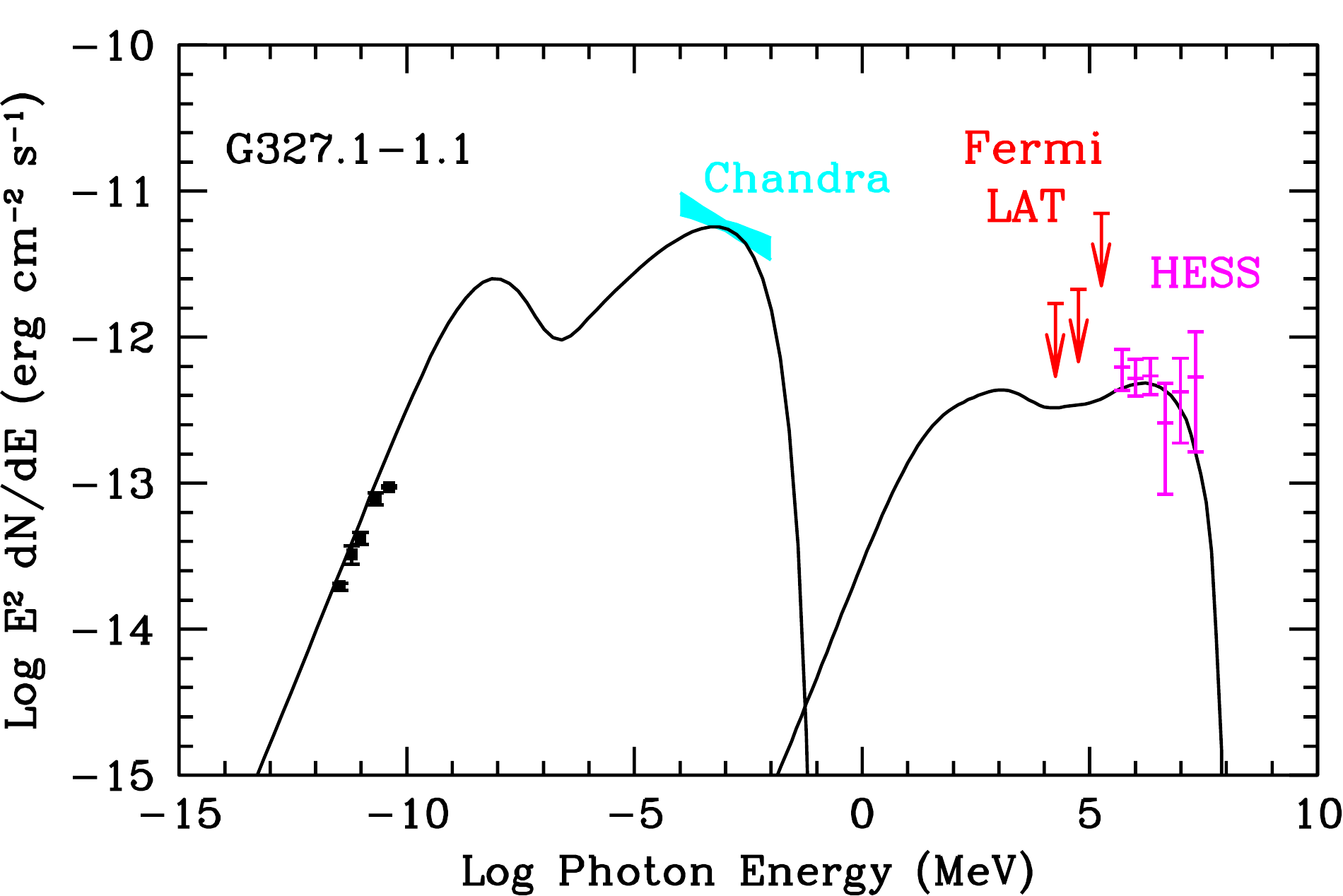}
\epsscale{1.1}\plotone{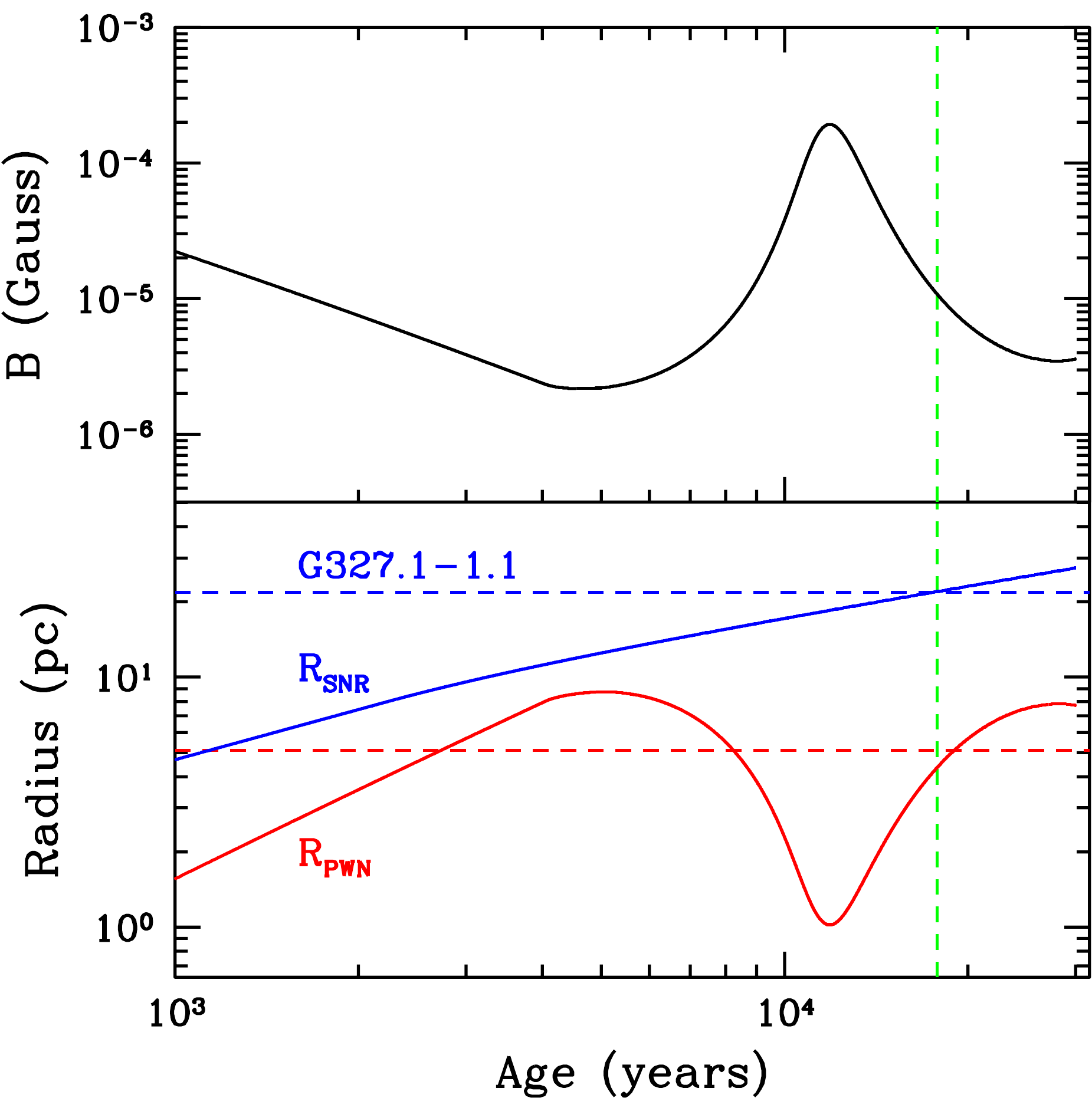}

\caption{\label{broadbandfig}Top: Model for emission from the evolved PWN in G327.1-1.1 (see Section~\ref{broadbandfig}) along with observed radio emission from the PWN in black \citep{ma15}, X-ray flux and spectral index in cyan (Section~\ref{spectra}), {\sl H.E.S.S.} $\gamma$-ray emission in magenta \citep{acero12} and {\sl Fermi} upper limits in red \citep{acero13}. Bottom: Time evolution of the PWN magnetic field (upper panel) and SNR and PWN radii (lower panel) for G327.1-1.1. The observed SNR and PWN radii are indicated by the blue and red horizontal dashed lines, respectively, and the age at which these radii are reached in the model is indicated by the green vertical dashed line.}
\end{figure}
%**************************************************

\section{Broadband Emission}\label{broadband}

To investigate the broadband emission from G327.1-1.1, we have used the semi-analytical model of \citet{gelfand09} for the evolution
of a PWN within the confines of an SNR. This one dimensional model simultaneously treats the magnetic, dynamical, and radiative evolution of the system, and given a set of input parameters, predicts properties such as the magnetic field, PWN and SNR sizes and expansion velocities, and the broadband spectrum of the PWN, as a function of the age of the system.
We use an ambient density of $0.12 {\rm\ cm}^{-2}$ based on our X-ray estimates for a distance
of 9~kpc, and evolve the system until the SNR radius matches the observed value of 22 pc. We use the results of our HD modeling for the explosion energy, ejecta mass, initial pulsar spin-down power, and pulsar spin-down timescale (see Table~\ref{properties}). We assume the injection of a magnetized electron/positron wind with a broken power law for the particle spectrum. The spectral index for the low energy component is fixed at 1.48 based the observed radio index \citep{ma15}, while that for the high energy component is fixed at 2.2 based on our measured X-ray spectral index for the emission in the compact source immediately surrounding the pulsar (see Section~\ref{spectra}). 

The resulting evolution of the SNR and PWN radii is shown in Figure~\ref{broadbandfig}. The model reaches the radius of G327.1-1.1 (indicated by
the blue dashed line) at an age of $\sim 17800$ yr. By this time, the PWN has been compressed by the SNR reverse shock, and has
begun to expand again, as seen from the evolution of the PWN radius. At the current age, the PWN radius is $\sim 5$~pc, in
good agreement with the size of the relic nebula in G327.1-1.1. 

The magnetic field in the PWN, shown in the upper panel of Figure~\ref{broadbandfig}, is evolved through the process of expansion and RS compression under the assumption that a constant fraction $\eta_B$ of the spin-down power is injected in the form of magnetic flux, with the
remaining fraction $\eta_e$ representing the injected particle flux. Radiative and adiabatic losses are calculated throughout the
evolution, and the final spectrum is obtained from the resulting electron population using the evolved magnetic field to calculate
the synchrotron radiation. Inverse-Compton emission is calculated for scattering of the electron population from photon fields
associated with the CMB, local starlight, and infrared emission from local dust \citep{strong00}. We find that an electron
spectrum comprising 99.7\% of the spin-down power at injection, with an energy break at $\sim 300$~GeV, produces a good representation
of the broadband spectrum, which is shown in Figure \ref{broadbandfig}. Here the X-ray spectrum (cyan) represents that of the entire PWN, as derived in Section \ref{spectra}, and the {\sl H.E.S.S.} spectrum (magenta) is that reported by \citet{acero12}. The {\sl Fermi} LAT upper limits (red) are taken from \citet{acero13}. The modeled magnetic field at the current SNR age is $10.8\:\mu$G, which based on the $\sim 300$~GeV intrinsic break in the electron spectrum, predicts a spectral break in the mid-infrared band.

We note that the treatment here is oversimplified in several ways as a result of the spherically-symmetric evolution into a constant
density in the semi-analytical model. This ignores the effects of the asymmetric crushing of the PWN, so the resulting magnetic field
evolution is not likely to reflect the more complicated behavior expected in G327.1-1.1. Despite these inadequacies, the model provides
a reasonable representation of the broadband spectrum, and of the PWN radius -- at least for the relic nebula. While the spectrum of the
high energy electrons corresponds to that derived from the emission close to the pulsar, the overall X-ray spectrum is produced by
electrons near the maximum energy, which are suffering from loss-induced steepening of the electron spectrum.

\section{Evolutionary State}\label{evol}

The modeling described in Sections~\ref{hdmodel} and \ref{broadband} indicates that the morphology and broadband spectrum of G327.1-1.1 is well described by the evolution of a composite SNR whose PWN has been crushed by the RS. A scenario in which the SN blast wave expands in non-uniform ISM density that decreases from west to east, and in which the pulsar is moving to the north with a velocity of 400 $\rm km\:s^{-1}$, matches the observed morphology of the G327.1-1.1 at radio and X-ray wavelength remarkably well. The observed structure of the SNR shell and relic PWN at radio wavelengths, as well as the morphology of the X-ray PWN and trail, are reproduced by the HD simulation at the age of $\sim$ 17000 yr, consistent with the age of G327-1.1 as derived from the Sedov model. Despite the oversimplification of spherical symmetry, the semi-analytic model of the evolution of G327.1-1.1 can match the broadband spectrum of the PWN emission from radio to $\gamma$-ray wavelengths, with input parameters consistent with the observed parameters and those derived by the HD simulation.

In earlier work, we speculated that the cometary X-ray PWN in G327.1-1.1 is likely not a bow shock, due to the fact that we observe emission ahead of the pulsar \citep{temim09}. In Section~\ref{compact}, we show that the compact source immediately surrounding the pulsar shows evidence for a torus-like structure embedded in the cometary PWN. This provides further evidence that the newly-formed PWN is not in the bow-shock stage of evolution, but that its cometary morphology is most likely a result of the passage of the RS that reaches the PWN from the NW direction (see Section~\ref{hdmodel}). In addition, MHD simulations of pulsar bow-shock and subsonic nebulae show that in a fully subsonic model,  the cometary nebula and X-ray trail are more confined and have  typical particles ages on the order of a few hundred years \citep{bucciantini05,swartz14}. This is consistent with our observations of the size and spectral steepening of the X-ray PWN in G327.1-1.1. 
We note that these same models predict a much smaller size for the termination shock radius than the size of the compact source that we attribute to a torus-like structure, suggesting that the structure that we are observing is a Crab-like torus that persists downstream of the termination shock, rather than a ring from the termination shock itself.

While the above scenario provides striking consistency with the observed spectral and spatial properties of G327.1-1.1, some outstanding questions still remain. The prong and arc structures are the most puzzling and difficult to explain, especially since they don't seem to originate at the pulsar itself. None of the HD simulations of the PWN crushing process produced non-thermal emission ahead of the pulsar's motion. One possibility is that these are structures associated with enhancements in the magnetic field. 
In Section~\ref{spectra}, we noted that the prong and arc structures are reminiscent of a funnel-shaped outflow aligned with the X-ray trail (i.e., the radio finger). The fact that this ``outflow" is oriented in the NW direction, the effective direction from which the RS first reached the pulsar and the PWN, may suggest an association with the RS/PWN interaction. Three-dimensional HD simulations of this system may provide additional clues into the origin of these unusual  structures.

\section{Conclusions}\label{conclusions}

In this paper, we analyzed 350 ks \textit{Chandra} observations of the composite SNR G327.1-1 and showed that the emission originates from a thermal shell with a temperature of $0.29^{+0.05}_{-0.04}$ keV, and a PWN with an average photon index of $2.15^{+0.05}_{-0.04}$. The \textit{Chandra} imaging reveals the complex morphology of the PWN; a compact source surrounding the presumed pulsar, embedded in a cometary X-ray nebula with a trail of emission  behind it. The deeper X-ray observation reveals non-thermal emission coincident with the radio PWN, with a steep photon index of $2.58^{+0.07}_{-0.10}$. There is also evidence for thermal emission with enhanced abundances in the PWN region. Morphological evidence suggests that it originates from SN ejecta mixed-in with the PWN material following a RS interaction. Unusual prong-like structures originate from the vicinity of the PWN and extend into large arcs. The structure is reminiscent of a funnel-shaped outflow towards the NW direction. The deconvolution of the compact source immediately surrounding the pulsar shows some evidence for a torus-like shape, which supports our conclusion from previous work that the PWN has not formed a bow shock and is expanding subsonically. The average absorbing column density for the SNR is approximately $2\times10^{22}\rm \: cm^{-2}$, with evidence of higher values in the western portion of the SNR.

HD simulations of a composite SNR expanding in a non-uniform density medium and hosting a rapidly moving pulsar reproduces the morphology of G327.1-1.1 exceptionally well. The simulation was constrained by the observationally determined SNR and PWN properties and evolved to an age of $\sim$17000 yr, the estimated age of G327.1-1.1 from the Sedov calculation. It shows that the morphology of G327.1-1.1 most strongly depends on the strength and orientation of the ISM density gradient, and the pulsar's velocity, spin-down luminosity, and  spin-down time frame. A scenario in which the ambient density gradient decreases from west to east, and the pulsar moves to the north with a velocity 400 $\rm km\:s^{-1}$, causes the RS to crush the PWN preferentially from the NW direction and produces a morphology that closely resembles the radio and X-ray observations of the SNR. The RS/PWN interaction scenario can also reproduce the broadband spectrum of the PWN at radio, X-ray, and $\gamma$-ray bands. Future modeling of similar systems whose physical properties can be constrained by observations promise to significantly improve our understanding of the evolution of composite SNRs.
 
\acknowledgments

This work was supported by NASA under the grant number GO2-13075A. P. S. acknowledge support from the NASA Contract
NAS8-03060.

\bibliographystyle{apj}

\end{document}